	\documentclass{jfm}
	\usepackage{graphicx}

	\shorttitle{ Entrainment  induced by waves at a density interface}
	\shortauthor{J. Herault et al.} 

	\title{    Entrainment  induced by waves at a density interface impinged by a turbulent jet}

	\author
	 {
	J. Herault \aff{1,2}
	  \corresp{\email{herault@irphe.univ-mrs.fr}},
	G. Facchini \aff{2},
	  \and 
	M. Le Bars\aff{2}
	  }
	
	\affiliation
	{
	\aff{1} 
	Institut de Radioprotection et de S\^uret\'e  Nucl\'eaire (IRSN), 13115 St Paul lez Durance, France.
	\aff{2}
	Aix Marseille Univ, CNRS, Centrale Marseille, IRPHE,  UMR 7342,
49 rue F. Joliot-Curie, 13013 Marseille, France
	}
	
\begin{document}
	
	\maketitle
	
	\begin{abstract}
{  Using water/salty-water laboratory experiments}, we   investigate  the mechanism of erosion by a turbulent jet   impinging  on  a density interface,  for moderate Reynolds  and  Froude numbers.   {  Contrary to previous models involving baroclinic instabilities}, we show that the entrainment is driven by interfacial  gravity waves, which break  and induce  mixing. The waves are generated by the turbulent fluctuations of the jet and   are amplified by a mechanism of {  wave-induced stress}.  Based on the physical observations, we introduce a scaling model which varies continuously from the  $Fr_i^3$ to the $Fr_i$ power law from small to large Froude numbers, in agreement with  {  some of the  } previous laboratory data.

	\end{abstract}
	

	
	\section{Introduction}

%
 	Turbulent mixing  in stratified flows is found in many geophysical and industrial situations and relies on a complex process where the kinetic energy is irreversibly converted into potential energy \citep{fernando1991turbulent,fernando1996some}. The present paper considers the case of a sharp density interface impinged    by a turbulent  round jet. The turbulent jet of light fluid (water) of density $\rho_1$  impinges on a volume of denser fluid (salty water) of density $\rho_2>\rho_1$, such that the  outflow of the jet is orientated in the gravity direction $\vec g$ and orthogonal to the density interface. The jet is driven by an upstream pump.  The salty water is progressively mixed with the fresh water until the gradients of density disappear. In  this process, two mechanisms can be identified: the   entrainment and the   mixing.  The entrainment corresponds to the advection  at the interface of dense fluid patches into the lighter fluid, where the flow is turbulent. The  mixing  consists in the thinning of the fluid patches by the turbulence until  the  spatial scale reaches the molecular diffusion scale, where both fluids are irreversibly mixed.  The turbulent entrainment is quantified by the entrainment rate \citep{baines1975entrainment}
	
\begin{equation}
E_i=\frac{ Q_e }{ \pi {b_i}^2  u_{i}}
\label{eq:scalinglaw}
\end{equation}	

	\noindent  where $Q_e$ refers to the volumetric flux entrained across the interface and  $u_{i}$ and $b_i$ correspond  to the vertical velocity and the width of the jet  at the interface. \cite{baines1975entrainment} suggested that for a turbulent flow, the entrainment rate $E_i$ depends only on   the interfacial Froude number,  where   the Froude number is  $Fr_i=u_{i}/\sqrt{b_i   g'}$ with $ g'=g (\rho_2-\rho_1) /\rho_1 $ the reduced gravity  acceleration. The Froude number characterizes the competition between the inertial force, which perturbs the interface, versus the restoring buoyancy force. Since \cite{baines1975entrainment}, the entrainment flux has been commonly expressed as a power-law function of $Fr_i$ with $E_i \propto Fr_i^n$ and $n \in[0,3]$. 
	
	The significant scatter in measurements of the entrainment rate has been summarized   by a recent review of \cite{shrinivas2014unconfined}  for   jet, plumes and fountains. They report the   entrainment rates  of different studies \citep{baines1975entrainment,kumagai1984turbulent,baines1993turbulent,lin2005entrainment} and  show  that at low Froude number, two distinct trends are observed. The measurements of \cite{linden1973interaction}, \cite{baines1993turbulent} and \cite{lin2005entrainment} for   fountains  and  \cite{baines1975entrainment} for   plumes follow  approximatively the scaling law $E_i=0.07 Fr_i^3$  for $0.3<Fr_i<1.4$. The measurements of \cite{kumagai1984turbulent} for a  plume   display  a distinct branch with  larger entrainment rates. \cite{kumagai1984turbulent} originally characterized the scaling law by $E_i \propto Fr_i^3$ but \cite{shrinivas2014unconfined} propose the scaling law $E_i=0.24 Fr_i^2$. \cite{cotel1997laboratory} have also studied the entrainment rate for an impinging jet. They observed two different behaviours: a linear scaling with $E \propto Fr_i$ for $0.3<Fr_i<1$, which decays to a constant for  $Fr_i<0.14$.  A universal law for the   entrainment rate as a function of the solely Froude number seems to be ruled out by the previous experimental investigations, which suggest  that other parameters must be taken into account. 

	\cite{breidenthal1992entrainment} and  \cite{cotel1996model} proposed to characterize the regimes with   the    Reynolds numbers $Re= u_i b_i /\nu$, where $\nu$ is the kinematic viscosity, in addition to the Froude number. \cite{baines1975entrainment}  showed that the entrainment rate does not depend on the Reynolds number providing the flow is turbulent.   However, \cite{shy1995mixing}   observed that the properties of the flows inside the impinged region change drastically when the Reynolds number is increased. For moderate Reynolds number ($10^3<Re<10^4$), the density interface in the impinged region remains sharp with a small mixed layer thickness but above $Re \simeq 10^4$,   the   thickness  of the density gradient increases by a factor $10$ :  it is  the so-called mixing transition.  \cite{shy1995mixing}  interpreted this transition as  a strong enhancement of baroclinic turbulence inside the dome. \cite{cotel1997laboratory} verified the presence of this transition, but the authors confirmed  the experiments of \cite{baines1975entrainment} by showing  that $E_i$ remains independent of the Reynolds number  for $Re \in[2200-12400]$.   The problem of  how the phenomenology   differs at the transition without modifying $E_i$ is still open. 
	
	 Describing the turbulent flow as a superposition of vortices, the models of erosion are commonly based on the mixing capacity of vortices, determined by their Froude number $Fr_\lambda=u_\lambda/\sqrt{ \lambda g'}$, with radius $ \lambda$ and velocity $u_\lambda$ \citep{linden1973interaction,shy1995mixing,cotel1997laboratory,cotel1996model}.  Only the vortices with  $Fr_\lambda>1$ are expected to contribute significantly to the entrainment in the impinging region. The classical model of engulfment of \cite{linden1973interaction} states that the eddies  rebounding on the interface  engulf  the dense fluid before being advected by the mean flow outside the dome.  \cite{linden1973interaction} proposes a law  $E_i \propto Fr_i^3$ based on scaling  arguments and kinetic and potential energy balance.  \cite{shy1995mixing} and \cite{cotel1997laboratory}  report  the presence of persistent baroclinic vortices localised around the periphery of the dome. The baroclinic vortices are generated by the incident small scales vortices advected by the turbulent jet. From these qualitative observations, \cite{shrinivas2014unconfined} suggest  a mechanism of  entrainment based on a stationary circulation driven by   like-sign vortices forced by baroclinic effects. For  the low Froude number regime, it leads to a scaling   $E_i \propto Fr_i^{2}$ in unconfined geometries.    In  confined geometries, \cite{shrinivas2015confined} suggest that the presence of large scale oscillations of the interface leads to a correction $Fr_i^3$, so that the entrainment rate becomes $E_i \sim  Fr_i^{2}+K Fr_i^{3}$ , with a constant $K$ vanishing in the unconfined limit. {  They show that  the confinement  effect may explain the   scatter of the data observed in their $(Fr_i,E_i)$ graphics, i.e. the existence of two branches at low Froude number.  }

	The model of \cite{shrinivas2015confined}  attempts to rationalise the different entrainment laws but it relies on the presence of vortices driven by baroclinic forces, even if no experimental study has demonstrated quantitatively  their existence.  The presence of the mixing transition suggests that the strength of the baroclinic vortices inside the dome depends on the Reynolds number of the turbulent jet. Turbulence in jets with large Reynolds numbers ($Re>10^4$) is expected to be developed and to display  small scales structures, more efficient to generate baroclinic \citep{shy1995mixing},  whereas jets with moderate Reynolds numbers ($10^3<Re<10^4$) remain dominated by large scale structures \citep{dimotakis2000mixing}.    These points raise  two questions:  can the vortices of a turbulent jet drive baroclinic eddies for Reynolds numbers below the mixing transition? What is the mechanism of entrainment at moderate Reynolds numbers if the baroclinic vortices are absent? These questions will be  addressed here.

In the present paper, we aim at  determining the mechanism of entrainment of a non-buoyant jet impinging a sharp density interface via velocity and density measurements for Reynolds numbers below the mixing transition. Our experimental set-up is presented in section 2 and 3. Our study holds for moderate $Fr_i$, i.e.   $Fr_i <1$, corresponding to a relatively small deformation of the interface, and for  moderate Reynolds numbers ($Re \simeq 2  \times 10^3$). Our measurements support a mechanism where the entrainment  is  driven by  interfacial  gravity-waves as shown in section 4. The waves are generated by incident vortices and amplified during their propagation  outside the impinged region. This amplification leads to   wave breaking associated with  local mixing.  Based on our measurements and on this  mechanism, we  introduce in section 5 a model  describing the continuous change of the scaling law  for turbulent entrainment from small  to large  Froude number.  
%
	 \section{Experimental set-up }
	
\subsection{Set-up and measurements}	
	
\begin{figure}
	  	\hbox{ \centerline{(a) \hspace{6cm} (b)}}
	    \centerline{ \includegraphics[width=11cm,height=6cm]{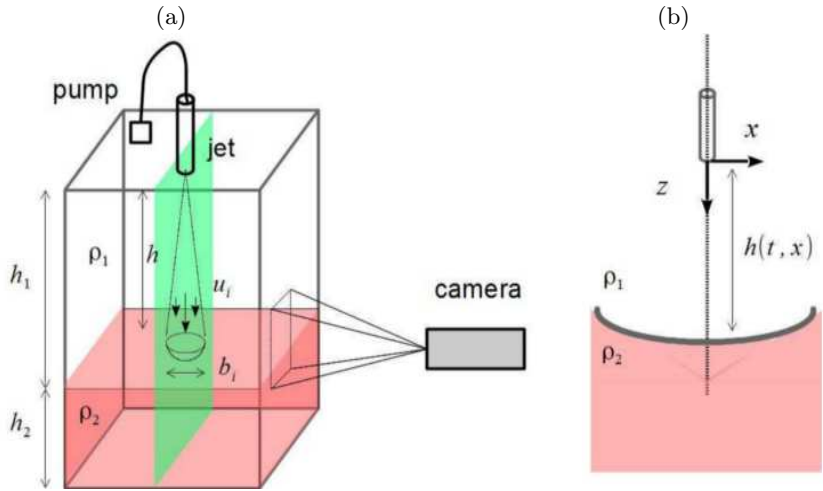}  }
	    \caption{(a) A schematic of the experimental  set-up   showing a rectangular tank filled with water ($h_1$) and a saline solution ($h_2$). The nozzle is located at a height $h_0=h(x=0,t=0)$ from the interface. (b) We define the reference  frame $(x,z)$ associated with the nozzle. The $z$ direction is descending.  \label{fig:setup}}
	 \end{figure}

	The set-up is sketched in figure \ref{fig:setup}(a). The tank is rectangular with  a square section of  width $L=30 $ cm and  height H$=50$ cm.  {  The tank is filled with one layer of light fluid of density $\rho_1$ and height $h_1=25$cm above a layer  of denser  fluid of density $\rho_2$ and height $h_2=14$cm (figure \ref{fig:setup}).	The densities are controlled by a density meter DMA 35 from Anton Paar.  The upper part of the tank is filled with water,  plus some ethanol  for the experiments indexed by M$1$, M$2$ and M$3$ in table \ref{tab:runs}}. The lower part of the tank is filled with a saline solution of density $\rho_2$.  The solution of ethanol in the light fluid allows us to match the optical index of the saline solution below. {   The viscosity of the mixture water plus ethanol is a non-linear function of the percentage of ethanol. The index matching is performed for density jump $\Delta \rho \simeq 0.03g/cm^3$.  The measurements are performed with  an initial density $\rho_1=0.987$, which is associated with a kinematic viscosity of $\nu=1.2 \times 10^{-6} m^2/s$ at $298$K \citep{khattab2012density}. The initial     viscosity is  $22\%$ larger than the   viscosity of the water but the mixing will progressively decrease this difference during the run.}   All the fluids are freshly prepared for each run and filtered (20 microns filter). The velocity of injection $U_{inj}$ at the round nozzle of radius $b_0 = 0.2$ cm is fixed for each experiment. The  pump for the injection is located in one of the upper  corner of the tank: the experiments are thus performed with a constant volume. {  Our measurements shows that the pumping system does not modify significantly the symmetries and  properties  of the mean velocity field.}

	\begin{table}
	 \begin{center}
	  \begin{tabular}{ccccccccc}
Run	& $U_{inj}$ $[m/s]$ &$ \Delta \rho [g/cm^3] $ & $h_0$ $[cm]$& $Fr_{inj}$	&	$\quad Re \quad $ 				&$\quad Fr_i \quad$ 	& $b_d$ $[cm]$& $\xi$  $[cm]$\\[3pt]
	\hline
	T1		&	1.15	&	$-$		&	$-$	&	$-$			&  $2.3 \times 10^3$	&	$-$	&	$-$	&	$-$		\\
	M1		&	1.03	&  0.031	&	$24$	&	41.8		&	$2.0  \times 10^3$	&	$0.56$	&	$4.0$	&	$2.0$ \\
	M2		&	1.15	&	0.031	&	$24$	&	46.6			&	$2.3 \times 10^3$	&	$0.63$	&	$3.4$	&	$0.8$	 \\
	M3		&	1.15	&	0.028	&	$24$	&	49.1		&	$2.3 \times 10^3$	&	$0.66$	&	$4.0$	&	$1.5$\\

	  \end{tabular}
	  \caption{		Experimental  parameters  for  the  4  conducted   experiments. }
 \label{tab:runs}
	 \end{center}
	\end{table}

 { 
	Two different experimental investigations are performed: the first one for the description of the turbulent jet without density variation, and  the second one for the investigation of the mixing mechanism. The runs are respectively indexed by ``T`` and ``M``(table \ref{tab:runs}). {   The physical  properties of the waves  (section \ref{sec:properties}) and  the features of the breaking  (section \ref{sec:breaking}) are determined from the measurements of the density field, run indexed by M2. The excitation  (section \ref{sec:generation}) and the amplification  (section \ref{sec:amplification}) of the waves are mostly characterized by the velocity measurements in runs indexed  M1 and M3. We have checked that the   runs display  the same features, like the signature of the wave period  in the measurements of density  M2 (figure \ref{fig:interface_ampl2}) and  in the measurements of velocity M1 (figure \ref{fig:uz_omega}(a))}. The measurements   start four minutes after the beginning of the erosion.  We performed both velocity  and density measurements using a digital camera (FASTCAM Mini, Photron) with a spatial resolution of $1024 \times 1280$  pixels.

	The velocity  field is measured in the experiments T1, M1 and M3, using  the  particle image velocimetry (PIV) process  \citep{meunier2003analysis}. The flow is seeded with tracer particles ($<10 \mu m$) and it is illuminated by a laser sheet crossing the vertical plane of the jet.    The dimensions of the interrogation windows are $15 \times 20$ cm${}^2$. The acquisition rate is performed at  $125$ fps   and $5000$ frames are acquired. The PIV algorithm is based on   a first $64 \times 64$ pixels and a second $32 \times 32$ pixels interrogation regions, and the resulting velocity field  is  a $64 \times 72$ vector field. The convergence of the mean field is insured  for an ensemble of frames larger than $4000$ snapshots. During the PIV measurements, the interface between the fluids is detected independently via a small amount of rhodamine in the dense fluid.  
}
	 {   	In  the experiment M2, the density is measured by a  planar laser-induced fluorescence technique (PLIF) with the same laser and camera set-up. The saline solution is mixed with rhodamine  B. This rhodamine has an absorption spectrum between 460 and 590 nm and its emission spectrum is 550-680 nm, well separated from the laser light. A high-pass filter (TIFFEN Filer, orange 21) is mounted on the camera with a cut off frequency of 550nm. The calibration has been performed with uniform fields of density.  The dye concentration has been adjusted to have a linear  relation between dye concentration and intensity. From this calibration process,  we have calculated the measured spatial intensity distribution, called also flat-field, which takes into account the biased of intensity due to the lighting inhomogeneity  or the camera response.  The density of rhodamine   is then calculated from the intensity field corrected by the flat-field. Here, we are only interested in  iso-density contours defining the interface.  We define the interface height $h(x,t)$  as the curve of iso-density $  \rho=(\rho_1+\rho_2)/2$.

	  The initial distance between the end of the nozzle and the interface is    $h(x=0,t=0)=24$ cm (runs $M$, see table \ref{tab:runs})  , i.e. $60$   times the diameter of the nozzle in order to obtain a    developed turbulent jet \citep{list1982turbulent}. To describe the flow, we use the frame  in the plane of symmetry of the  jet   (figure \ref{fig:setup}(b)) with the cartesian coordinate system $(x,z)$. The  $x$-axis is aligned with the radial direction and the $z$-axis corresponds to the axis of symmetry of the jet. The origin is located at  the nozzle outlet. Due to the axisymmetry of the configuration, we assume that the time-averaged quantity of the azimuthal component of the velocity fields, here in the $y$ direction, and the derivative of the average quantities with  respect to $y$, are  equal to zero. The measurements performed in the vicinity of the impinged region are expressed as a function of the radial distance $x$ and the distance from the interface $h_0-z$, where $h_0=\langle h(x=0 ) \rangle$ is the time average distance from the nozzle  to  the interface along the axis of the jet.
 
}

	\subsection{Interfacial dimensionless parameters}	
\label{sec:dim_para}

	In  table \ref{tab:runs}, we report the Froude numbers and the Reynolds numbers associated with the jet for each run:  $Fr_{inj}=U_{inj}/\sqrt{b_0   g'}$ and $Re=U_{inj} b_0/ \nu$. The Reynolds number is expected  to  be constant in the far field of the jet \citep{list1982turbulent}. The Reynolds numbers are between  $2 \times 10^3$ and  $2.3 \times 10^3$, which implies that our experiments are below the mixing transition \citep{shy1995mixing}, where the mixing layer thickness  is small.  We  investigate  the properties of the solely turbulent jet configuration in the next section.  {   From the Froude numbers $Fr_{inj}$, we calculate a relevant Froude number at the interface $Fr_{i}$ by interpolating the typical length $b_i$ and velocity $u_i$ at the interface. We use the classical law of evolution of the width of the jet $b$ and the axial velocity component $u_m$   with a distance $z$ from the nozzle \citep{fischer} }
	
	\begin{equation}
	b(z )=  \frac{d b}{d z}( z -h_s) \quad \hbox{and} \quad u_m(z )= U_{inj} \Lambda \frac{ 2 b_0}{   (z -h_s)}
	\label{def:ub_turb}
 	\end{equation}
	
	\noindent  with  $(d b/ dz)$  the spreading rate,  $\Lambda$ the velocity decay constant  and $h_s$ the virtual origin of the jet reported in   table \ref{tab:turb_prop}. We use  the top-hat definition for the typical length $b_i$ and velocity $u_i$. The top-hat definition  for a gaussian profile allows us to define $u_i=u_m(h ) /2$ and $b_i=b(h ) $ with $h$ the position of the interface.  {  The recent numerical experiment of  \cite{ezhova} supports the use of   equations (\ref{def:ub_turb}) showing that  the axial velocity component $u_m$ (Fig.6(a)) is modified only in the near field of the interface. }	The Froude number at the interface

\begin{equation}
Fr_i=\frac{u_i}{\sqrt{b_i g'}}
\end{equation}	

\noindent  writes  as a function of the position of the interface $h$ 
	
	\begin{equation}
	Fr_i=Fr_{inj} \frac{\Lambda}{(d b/ dz)^{1/2}} \left(\frac{ b_0}{h-h_s}\right)^{3/2}
	\label{def:Fr}
	\end{equation}
	
	\noindent Thus the Froude number decreases like $Fr_i \sim h^{-3/2}$. {  Our study is performed during a limited duration after the initial transient and, it is reasonable to consider that the Froude number has not varied significantly since the beginning of the experiment. }

%
	\begin{table}
	 \begin{center}
	  \begin{tabular}{cccccccccccccccccc}
						&	$b_0$[cm]	& 	$Re$			&	$z/(2b_0)$	&	$h_s/ (2 b_0) $ & $(d b/d z)$	& $ \Lambda$ \\
						\hline
	T1					&	$0.2$		&	$2.3 \times 10^3$	&	$40-85$	&	$2.29$		& $0.096$  &$5.51$			\\ \\
	\cite{hussein1994velocity}& $1.27$		&	 $47.7 \times 10^3$	&  $16-96$	&  $4$			& $0.093$& $5.9$  	\\
			 	\hline
	   \end{tabular}
	     {\caption{Parameters of the turbulent jet compared to the measurements of Hussein \textit{et al.} (1994). }}
	      \label{tab:turb_prop} 
	 \end{center}
	\end{table}

%
	

	 \section{Kinematics of the jet}
	
%

	\subsection{Jet without stratification}

	  \begin{figure} 
	  	\hbox{ \centerline{(a) \hspace{6cm} (b)}}
	      \includegraphics[width=6.5cm,height=5cm]{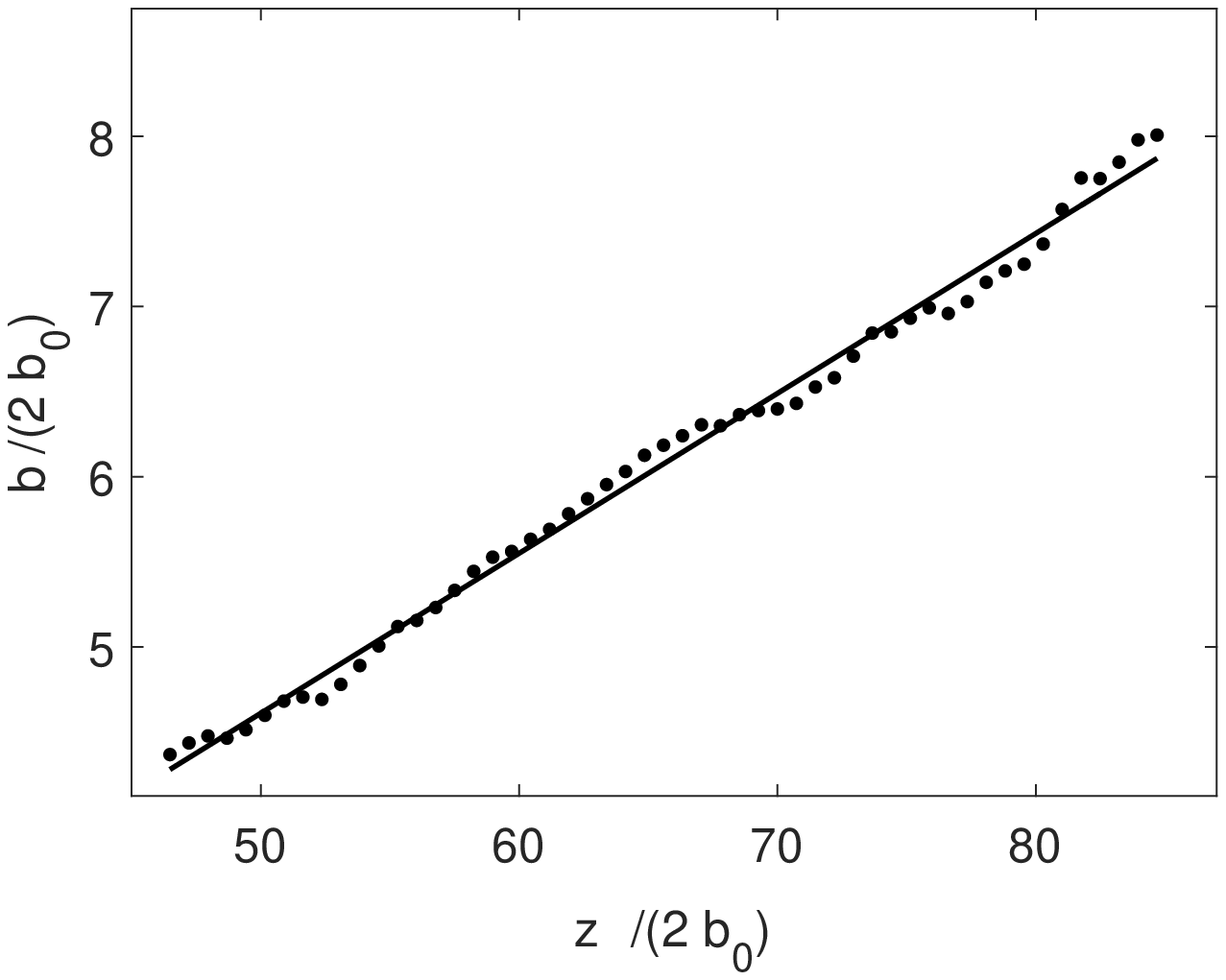} 
	    \includegraphics[width=6.5cm,height=5cm]{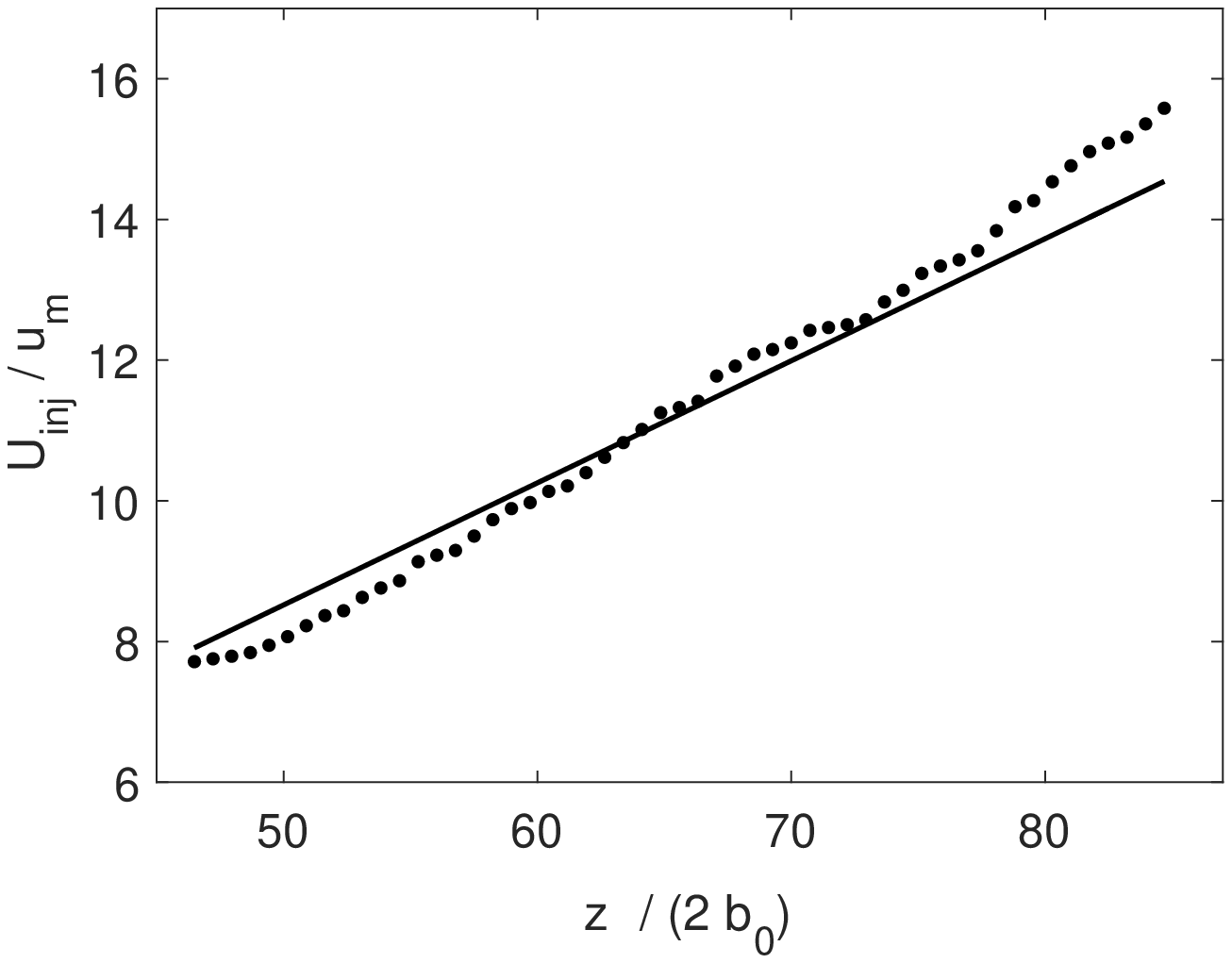} 
	    \caption{(a) Radius $b$   of the jet and  (b)  inverse of the axial velocity $u_m$ as a function of the distance from the nozzle $z$ rescaled by the diameter of the nozzle $2 b_0$.    \label{fig:jet_sans}}
	 \end{figure}
	 
%
	
	{  We first study the jet in the absence of stratification for a  Reynolds number  }$Re=2.3 \times 10^3$ (T1 in table \ref{tab:runs}) in order to validate the use of equation (\ref{def:ub_turb}). The measurements are performed for $z$  between $80 b_0$ and $170 b_0$. The local radius of the jet $b$  divided by the diameter $2 b_0$ is reported as a function of $z/(2 b_0)$ in figure \ref{fig:jet_sans}(a). The radius $b$ corresponds to the ordinate $x$ where the average vertical velocity corresponds to the half of the axial velocity, i.e. $\langle u_z \rangle(x,z)=u_m(z)/2$,  for a given  $z$.  The radius increases linearly and the fitted parameters $(db/dz,h_s)$ (equation (\ref{def:ub_turb})) are reported in   table 2. The ratio $U_{inj}/u_m(z)$ (figure \ref{fig:jet_sans}(b)) increases also linearly and the corresponding parameter $\Lambda$ is reported in   table 2. The coefficient $\Lambda$ is fitted by taking $u_m(z)$ as a function of $z-h_s$ with $h_s$ obtained from the data of figure \ref{fig:jet_sans}(a). The measured  coefficients  $(h_s,db/dz,\Lambda)$ are compared to the results of \cite{hussein1994velocity} for $Re= 47.7 \times 10^3$, based on the radius of the nozzle.   Our measurements are in good agreement, regarding $db/dz$ and $\Lambda$, which are related to the physics of the mixing by the turbulent jet: this validates the use of Taylor entrainment constant. The virtual origin is different but it depends strongly on the range of fitting  $z/b_0$, as demonstrated by \cite{wygnanski1969some}. 
 

	\subsection{Jet with  stratification}
	\label{sec:jetstratif}
	
\begin{figure}   
\centerline{	\includegraphics[width=8cm,height=6cm]{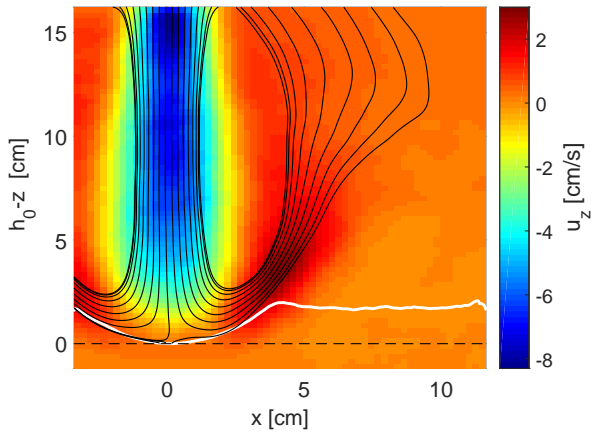}}
	\caption{ Streamlines (black curves) of the mean flow, superimposed on the vertical component of the mean velocity field $\langle u_z 		\rangle$ (cm/s) (colorbar). The white curve corresponds to the time-averaged location of the interface separating both fluids. The black dotted line corresponds to the ordinate $h_0-z=0$.  \label{fig:velocity_jet1}}
	\end{figure}

	We now focus on the velocity field  of a jet impinging on a sharp density interface. The results presented here correspond  to the run M1 (table \ref{tab:runs}). The    vertical component $\langle u_z \rangle$ of the time-averaged velocity field and its streamlines are reported on    figure \ref{fig:velocity_jet1}. {  The operator $\langle \cdot  \rangle$ corresponds to the time-averaged quantities calculated on a duration $T_e=40$s corresponding approximatively to $100$ advection time scales $b_i/u_i$}. The white curve represents the time-averaged location of the interface {  determined by the presence of fluorescent rhodamine}. The vertical axis at $x=0$ corresponds to the jet axis and the height $h_0-z=0$ is associated with the bottom of the time-averaged interface (black dashed line). 

{   We have reported in table \ref{tab:runs}, the depth and the radius of the  time-averaged  dome generated by the impact of the jet on the stratified interface for M1, M2 and M3. The radius $b_d$ is the radial distance   between the center of the dome (almost parabolic) and the end of the dome, where the interface becomes flat. The depth $\xi$ corresponds to the difference of altitude between these two extremities. For M1,} the dome is characterized by a radius    $b_d=4$cm and a depth $\xi=2$cm. The abscissa $x$ and the height $z$ will be rescaled respectively by $b_d$ and $\xi$ in the rest of the paper. The radius of the jet at the interface $b_i$, equal to $2.2$cm, corresponds approximately to the width of the jet entering in the dome and it is approximatively two times smaller than the radius of the dome.

	 	\begin{figure}  
	 	 	\hbox{ \centerline{(a) \hspace{6cm} (b)}}
	 	 	\includegraphics[width=6cm,height=5cm]{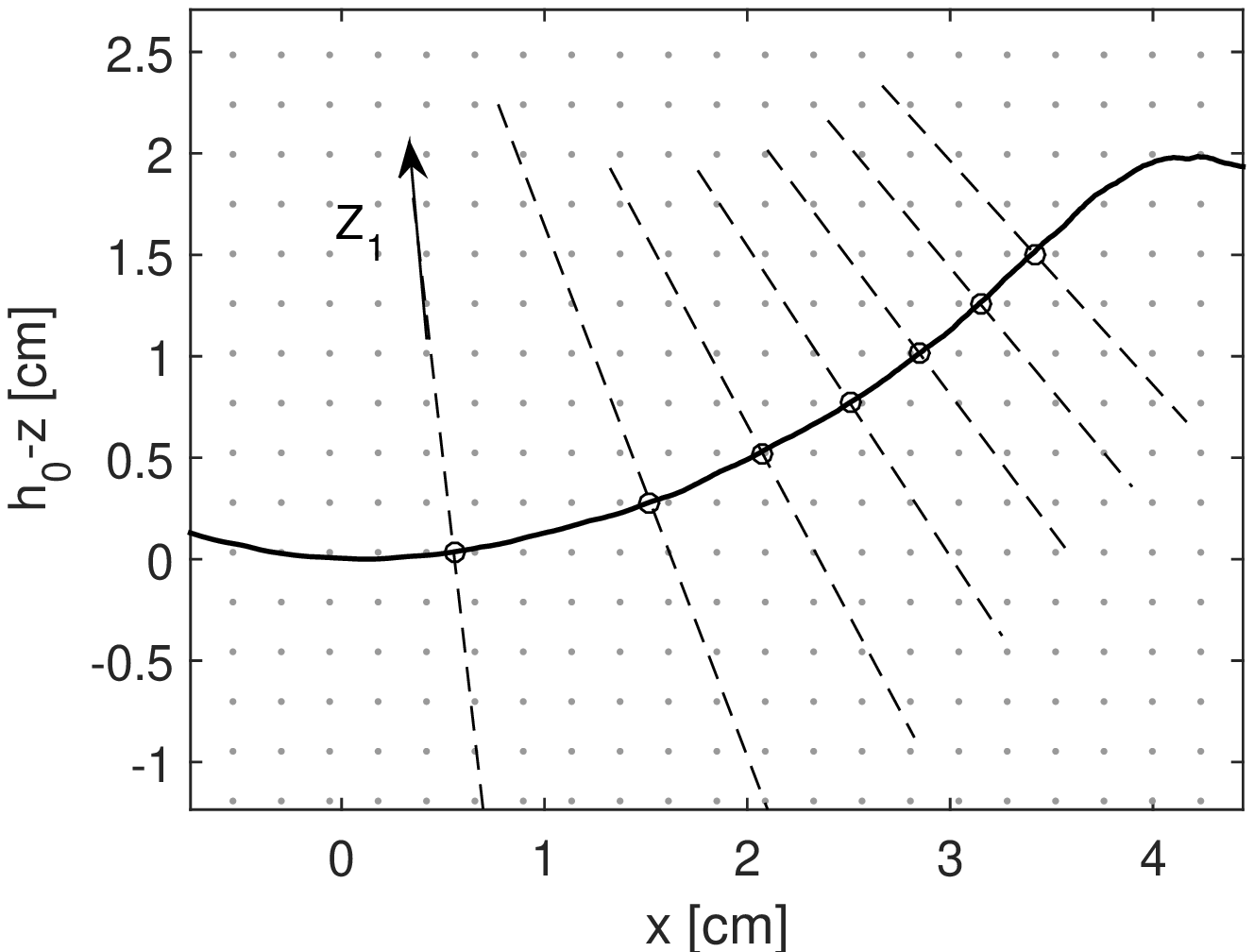}  
	 	 		\includegraphics[width=7cm,height=5cm]{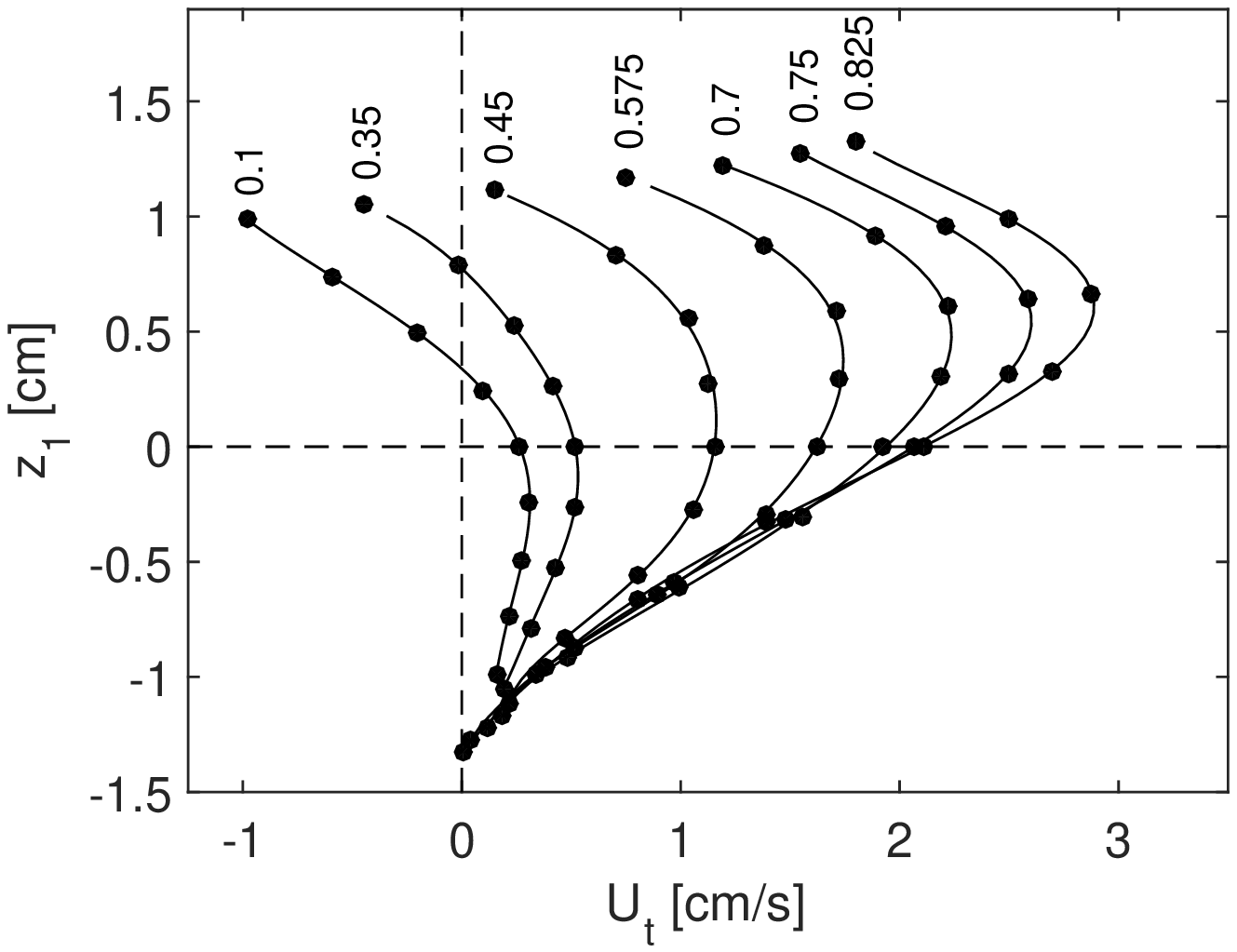}

	\caption{(a) Location of the points  on the interface with  the  normal line (dashed line) where the tangential velocity $U_t$  is calculated. (b) The profiles of $U_t$ are calculated at the abscissa  $x=0.1,0.35,0.475,0.575,0.7,0.775,0.825$ times the radius of the dome $b_d$. The variable $z_1$ is the distance from the interface following the dashed curves in (a).  \label{fig:velocity_jet11}}
	\end{figure}

	A remarkable property of the flow inside the dome is the convergence of the streamlines   leaving the dome close to the interface. This contraction corresponds to a local acceleration of the flow. To study the flow field inside the dome, we introduce  the velocity component $U_t$, which is tangential to the mean interface given by the function $\langle h (x,t)\rangle$. For each abscissa  $x$, we  calculate  the angle of inclination of the interface  given by $\theta(x) = \arctan( d \langle h \rangle/dx)$ and we define the velocity by $U_t(x,z)=\sin(\theta) \langle u_z\rangle+\cos(\theta) \langle u_x \rangle$. The profiles of $U_t$  are reported in figure \ref{fig:velocity_jet11}(b) for different   abscissae shown in figure \ref{fig:velocity_jet11}(a)(from left to right) starting  at $x=0.4$cm ($0.1 b_d$) and finishing at $3.3$cm ($0.825 b_d$). They are plotted as a function of the ordinate $z_1$  following  the normal to the interface (dashed lines in figure \ref{fig:velocity_jet11}(a)) with  the interface at $z_1=0$. The curves correspond to the polynomial interpolation of the profiles. The profiles confirm that the flow accelerates along the interface.

\begin{figure} 
	\hbox{ \centerline{(a) \hspace{6cm} (b)}}
	\includegraphics[width=6cm,height=5cm]{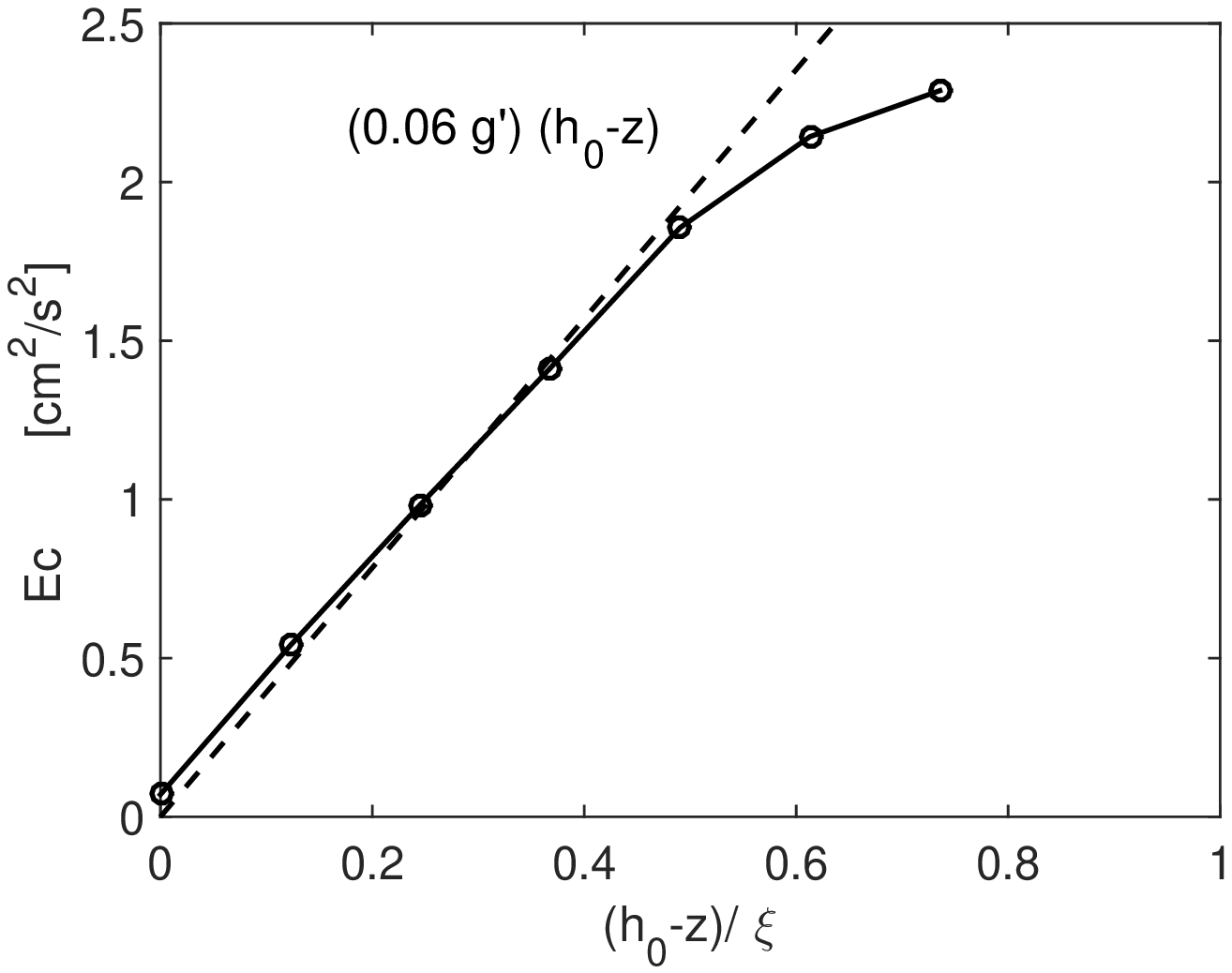} 
	\includegraphics[width=8cm,height=5cm]{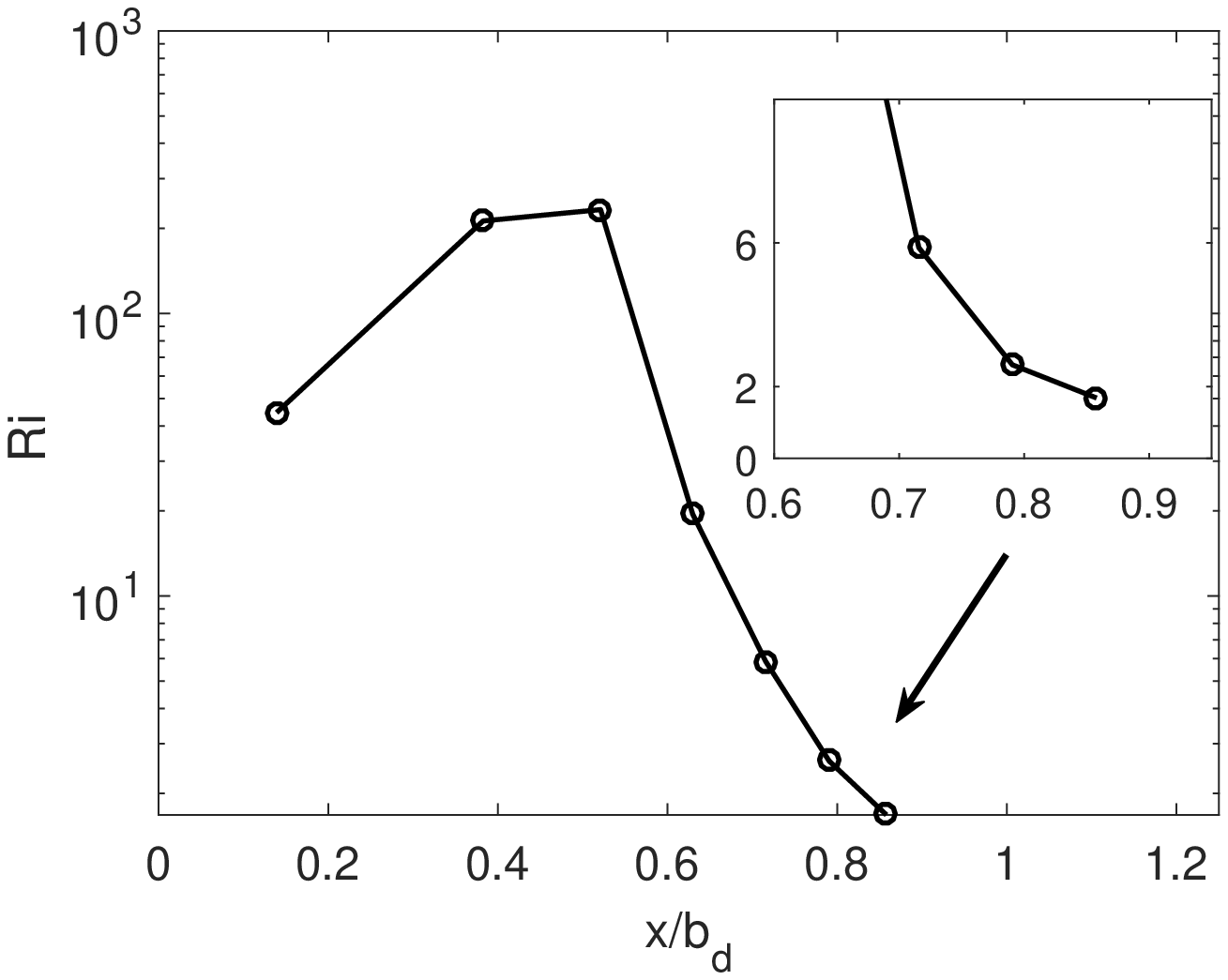}  
	\caption{(a) Kinetic energy $E_c$  as a function of the height along the interface. (b)	 Richardson number  as a function of $x/b_d$  with $Ri=  (g'/b_d )/s_0^2  $ and $s_0=(\partial_{z_1} U_t)_{z_1=0}$.    \label{fig:velocity_jet2}}
	\end{figure}

A balance between potential energy and kinetic energy explains the acceleration characterized by $\partial_x U_t>0$. When the jet impacts the interface, the kinetic energy is converted partially into potential energy. When the flow leaves the dome, a part of the potential energy is reconverted into kinetic energy. We have calculated the kinetic energy  $E_c= (\langle u_x \rangle^2+\langle u_z ^2\rangle) /2$  as a function of the height $z$ along the interface (figure \ref{fig:velocity_jet2}(a)).  The kinetic energy  increases initially linearly with the height $z$ until $z\simeq 0.6 \xi$, which corresponds to the abscissae $x \simeq 0.8 b_d$.  The time-averaged interface follows the  streamline of the time-averaged flow passing close to the stagnation point at $(x,z) =(0,0)$. {    \cite{shrinivas2014unconfined} (equation (3.9))  suggest that despite the presence of turbulence,  the Bernoulli equation can be applied along the streamline following the interface. By assuming a pressure balance at the interface and a vanishing velocity below the interface, they obtain  $E_c=  g' z$. However, our measurements show  that the kinetic energy follows $E_c=  ( 0.06  g' z)$: only six percent of the potential energy  is transferred into kinetic energy}. The application of the Bernoulli equation breaks because the velocity does not vanish rapidly below the interface (figure \ref{fig:velocity_jet11}(b)) and  it considers  neither the process of entrainment nor the turbulent fluctuations, which must consume an important part of the potential energy. Outside the dome,  the streamlines are directed upward and reconnect the jet flow drawing a large recirculation cell. This recirculation is the combination of the ballistic reflection of the jet and the  lateral entrainment of the jet.

	\section{Investigation of the  mechanism of erosion}
\label{sec:mecha}

	\subsection{General description}

\begin{figure}  
	\hbox{ \centerline{(a) \hspace{6cm} (b)}}
	
	\centerline{  \includegraphics[width=6.5cm,height=6cm]{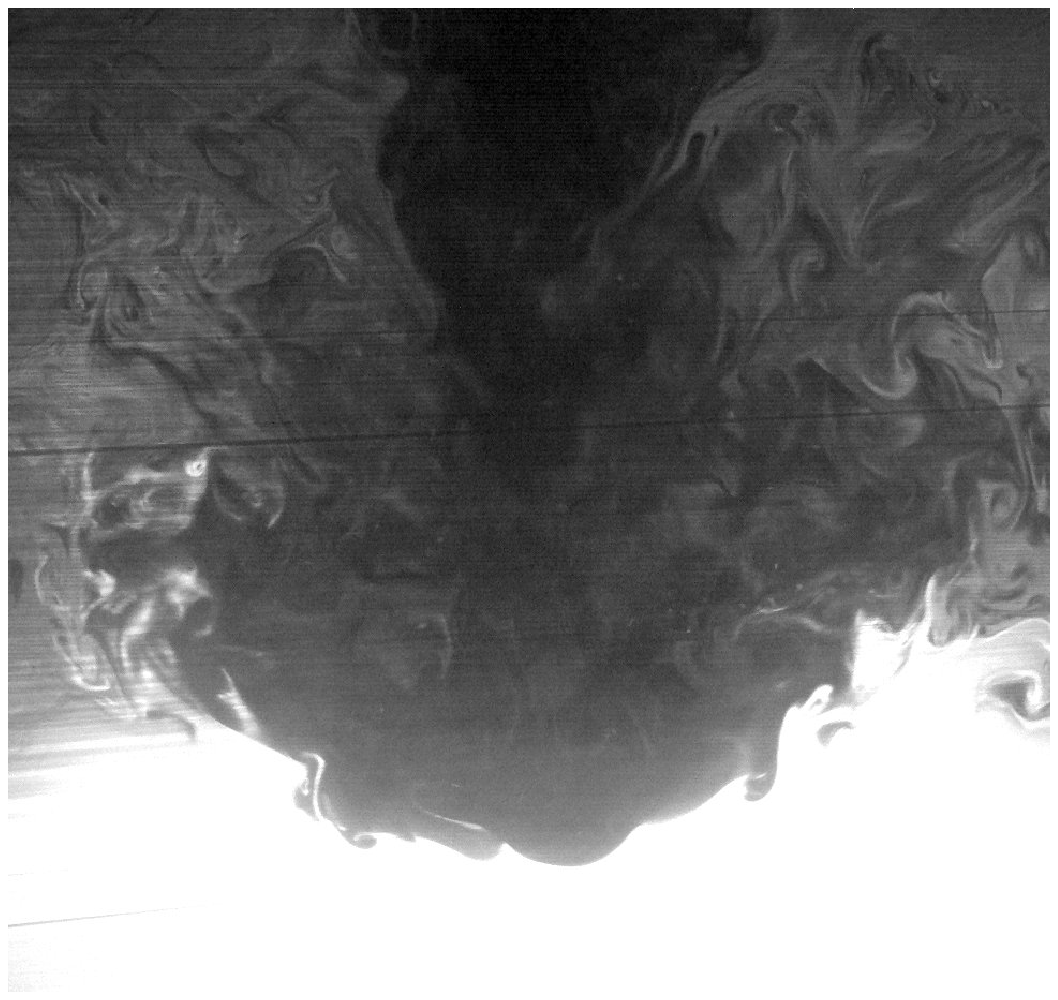}  
	 \includegraphics[width=6.5cm,height=6cm]{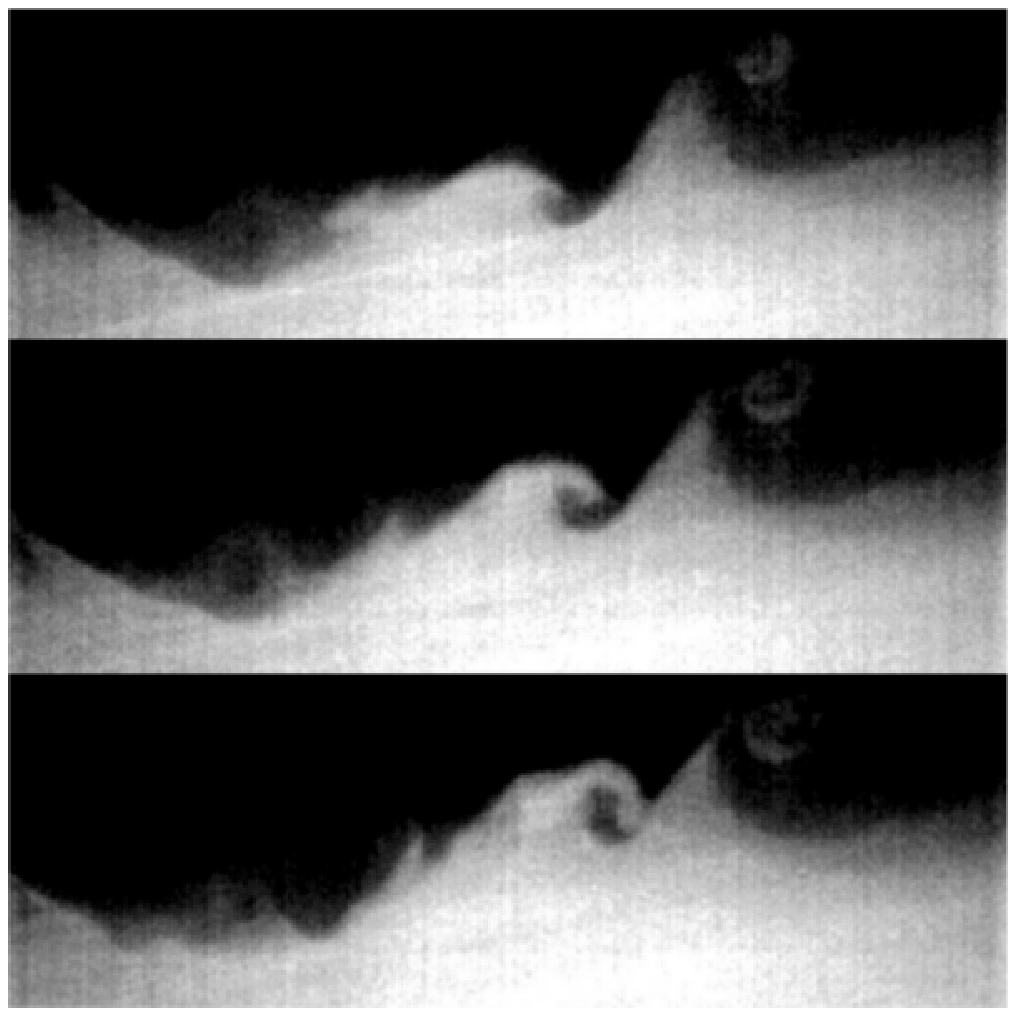} }
	    \caption{(a) Mixing process in the vicinity of the interface ($15 \times 20$cm${^2}$) and (b) of the dome head ($15 \times 8$cm ${^2}$). The dense fluid  has been dyed by rhodamine and corresponds to the bright region. See online supplementary material (movie1) for the movie corresponding to  figure (a). \label{fig:image_erosion1}}
	  
	 \end{figure}

	 We first draw a global picture of the mechanism of erosion before demonstrating each element in the following paragraphs. When the jet impinges on the dense fluid, the interface  is  convoluted by the turbulent fluctuations of the jet. These perturbations are generated by   eddies coming from the jet. These vortices excite {  interfacial gravity} waves  propagating outward the impinged region (figures  \ref{fig:image_erosion1}). During the propagation, the height of the waves increases until they breakdown close to the border of the dome. This process is illustrated on the three successive pictures of the interface of   figure \ref{fig:image_erosion1}(b), with  waves propagating from left to right. The amplification of the waves is caused by an energy transfer from the mean flow via a mechanism, which could be similar to the one described by \cite{miles1960generation} in shear flows. The breaking of the waves induces a strong  mixing due to the wrapping of  filaments. The heavy fluid is then transported in the bulk flow where the turbulence of the jet mixes both  fluids (figure  \ref{fig:image_erosion1}(a)). {   We claim } that the transport and the mixing of the dense fluid by these waves is   the  main source of erosion  at low Froude number and moderate Reynolds numbers. { 	 The following sections detail successively the former points.}

%
%

	\subsection{Wave properties}
\label{sec:properties}	

{   In this section, we  characterize the properties of the perturbations near the interface. We observe that the interface remains sharp in the impinged region, as expected for $Re<10^4$ \citep{shy1995mixing} with a small    diffusion layer $\delta_H$. The height of $\delta_H$  corresponds to few percent of the dome depth, i.e. few millimeters. The stratification in this layer could support internal gravity waves with typical frequency  $N=\sqrt{g' /\delta_H}$ with $\Delta \rho/\delta_H$ the associated density gradient. The   frequency of internal gravity waves would be  $ N\simeq 8$Hz for $\delta_H \simeq 4$mm.   As we will see further, the typical frequency of the waves propagating inside the dome  varies in the range $0.3-0.4$ Hz, a frequency range corresponding to interfacial gravity waves. Hence, the dynamics of the perturbation inside the dome is   controlled by interfacial gravity wave rather than internal gravity wave.

 The spatio-temporal properties of the waves is studied by using  the information contained in the dynamics of the interface. The perturbation of the interface is illustrated by an instantaneous relative density field from the measurement M2  (figure \ref{fig:interface_ampl}). The relative density is defined by $\hat{\rho}= (\rho-\rho_1)/(\rho_2-\rho_1)$ varying between $0$ and $1$, with $\rho$ the effective density given by the LIF measurements. The interface position defined by  $h(x,t)$  (black curve) corresponds to the iso-density line $  \tilde{\rho} =0.5$. The time-averaged height $\langle h \rangle (x)$ is represented by a black dotted curve following a quadratic law $\langle h\rangle \sim x^2$ in the impinged region.

 \begin{figure}  
	
	    \centerline{  \includegraphics[width=12cm,height=6cm]{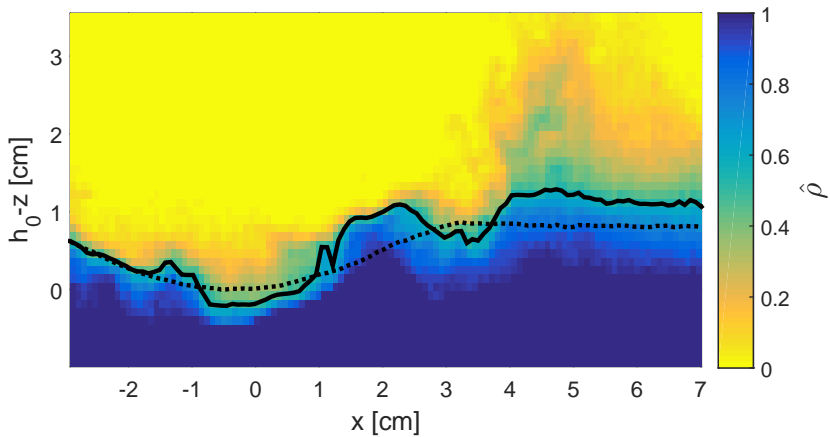}
	       } 
	    	
	    \caption{Rescaled density field $\hat \rho$  (colour level) from $\hat \rho=0$ for $\rho=\rho_1$ (fresh water)   to $\hat \rho=1$  for $\rho=\rho_2$ (initial salty water). The heavy black curve corresponds to the height $h(x,t)$ where $\hat \rho=0.5$. The dashed curve represents the  time-averaged height $\langle h \rangle$.    \label{fig:interface_ampl}}
	 \end{figure}

From the height $h(x,t)$, we introduce the height perturbation $\eta(x,t)=h(x,t)-\langle h\rangle $ and we perform a two-dimensional spectrum of the spatio-temporal evolution of $\eta(x,t)$. It consists in projecting the spatio-temporal field $\eta(x,t)$ inside the dome, i.e $\vert x \vert<b_d$, on the 2D Fourier space $(k,F)$.  We restrict the spatial domain  to the dome region, where the waves are generated and amplified.   We report in figure \ref{fig:spectrel2}, {  the two-dimensional power-spectrum  with  frequencies below $0.6$ Hz  and wavenumbers smaller or equal to $2 k_0$, with $k_0=2\pi/b_d$ (run M2, table \ref{tab:runs}). This bandwidth   contains $89$ percent of the total energy}.  The temporal dynamic of the height $\eta$ is mostly modulated by two time-spectral components:  slow standing modes  and  interfacial  gravity waves. The slow mode is characterized by frequencies in the range $F \in [0.05-0.15]Hz$ with a peak  in the mode with   wavelength $k=0.5 k_0$.  The waves are characterized by     wavenumbers between $0.5 k_0$ and $k_0$, i.e wavelengths between $b_d$ and $2 b_d$,   and   frequencies  in the range $F \in [0.3-0.4]Hz$, i.e a typical period  with $T=2.9 \pm 0.4$s. The energy in the waves is approximatively two times larger than the energy in  the slow modes. The precise characterization of the wavelength is difficult because the wavelength  is comparable with the length of propagation.   The  dispersion relation of interfacial  waves for irrotationnal flow gives a period of  $2 \pi  (k_0 g At)^{-0.5}  \simeq 1.85s$, a value comparable to the measured one. The difference may be explained by the presence of an inclined interface  of propagation or by  the   Doppler shift associated with a mean non-uniform flow.

	 }


%
 	 By using a bandpass filter containing the  frequency range  $[0.3-0.4]$Hz, we are able to  extract the propagation of the interfacial  waves. The spatio-temporal diagram of the bandpass filtered $\eta$   is reported on    figure  \ref{fig:interface_ampl2}(a), with the abscissae rescaled by $b_d$ and the time $t$ by the period of the waves $T$. The amplitude $\eta$ of the waves  is divided by the depth of the dome $\xi$. {  The spatio-temporal diagram displays  wave patterns  with a half-wavelength   comparable to the dome radius}. {   The  wave dynamic  is complex but we observe that most of the waves propagates outward  the dome, as illustrated by the  figure  \ref{fig:interface_ampl2}(b).} From the bandpass filtered elevation $\eta(x,t)$, we perform a lagrangian tracking of the waves, which consists in following the wave envelop. We   report  some trajectories $x(t)$  on    figure \ref{fig:onde_vitesse_ampl}(a) and calculate the phase velocity $c ={\dot x}(t)$. The velocities being not constant, we calculate the probability density function $p(c)$ of $c$ for all the trajectories (insert of figure \ref{fig:onde_vitesse_ampl}(a)). The distribution exhibits a maximum at $c_p=2   cm/s$, which will be taken as the reference phase velocity at the interface. The grey  dashed curve on  figure \ref{fig:onde_vitesse_ampl}(a) corresponds to a trajectory with $x=c_p t+x_0$.

	  \begin{figure} 
	    \centerline{  
	     \includegraphics[width=9cm,height=6cm]{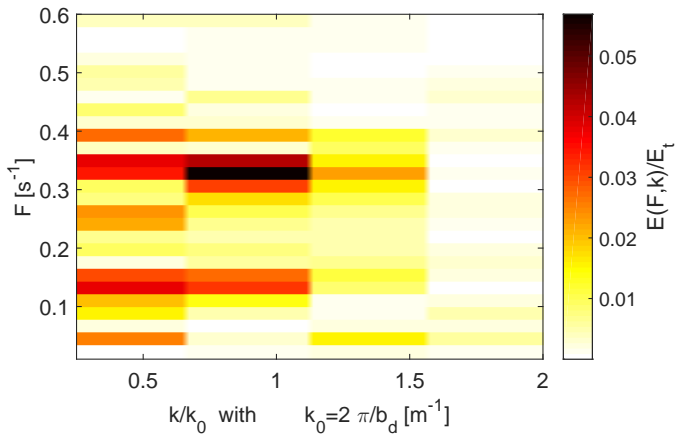}
	   } 
	    	
	    \caption{Two-dimensional power-spectrum $E(k,F)$ of the elevation $\eta(x,t)$.  The wave number $k$ is rescaled by $k_0=2 \pi/b_d$ and the density of energy is rescaled by the total energy $E_t$.     \label{fig:spectrel2}}
	 \end{figure}

 	  \begin{figure} 
 	\hbox{ \centerline{(a) \hspace{6cm} (b)}}
	
	    \centerline{  \includegraphics[width=7.25cm,height=6cm]{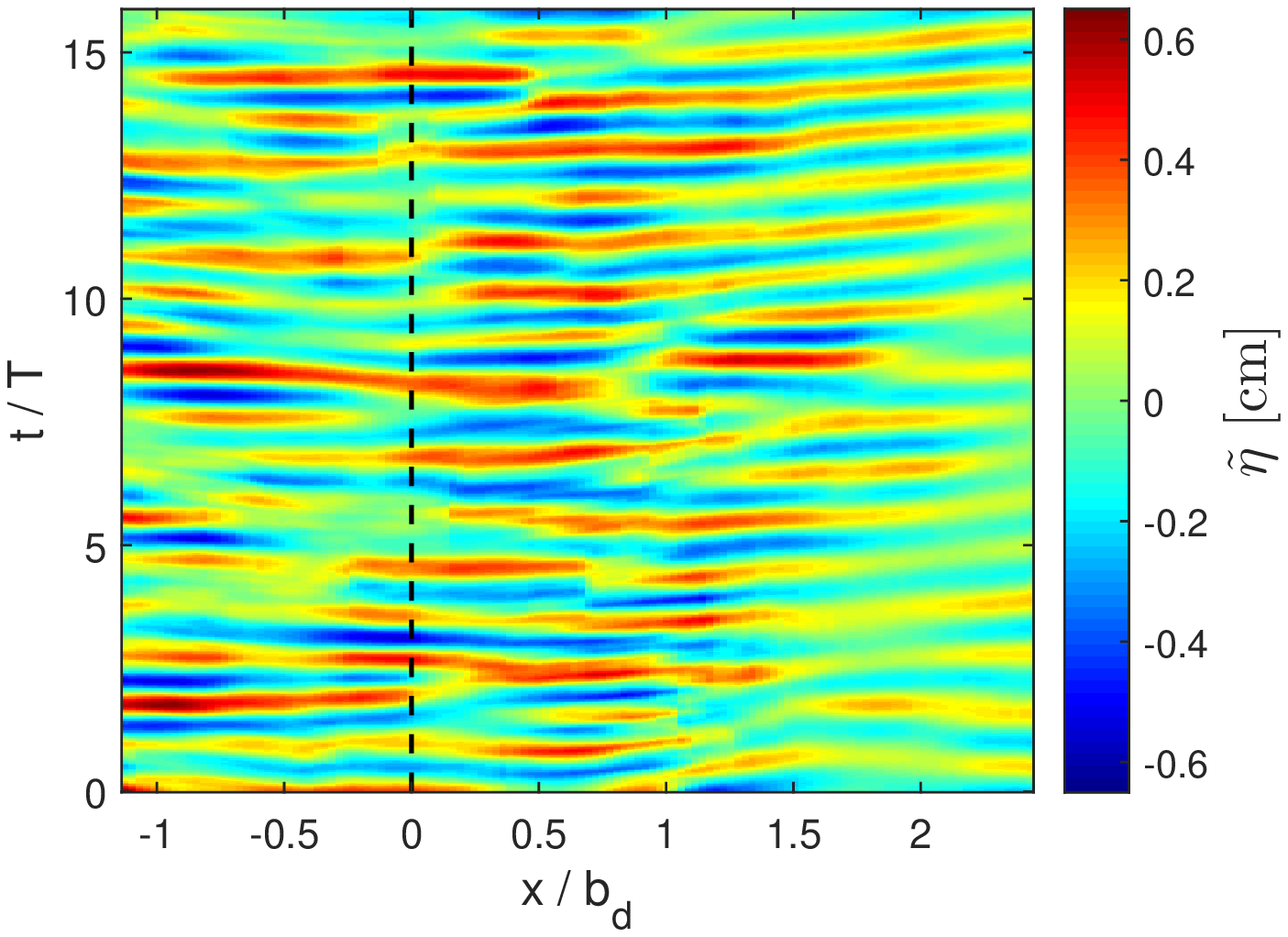}
	     \includegraphics[width=7.25cm,height=6cm]{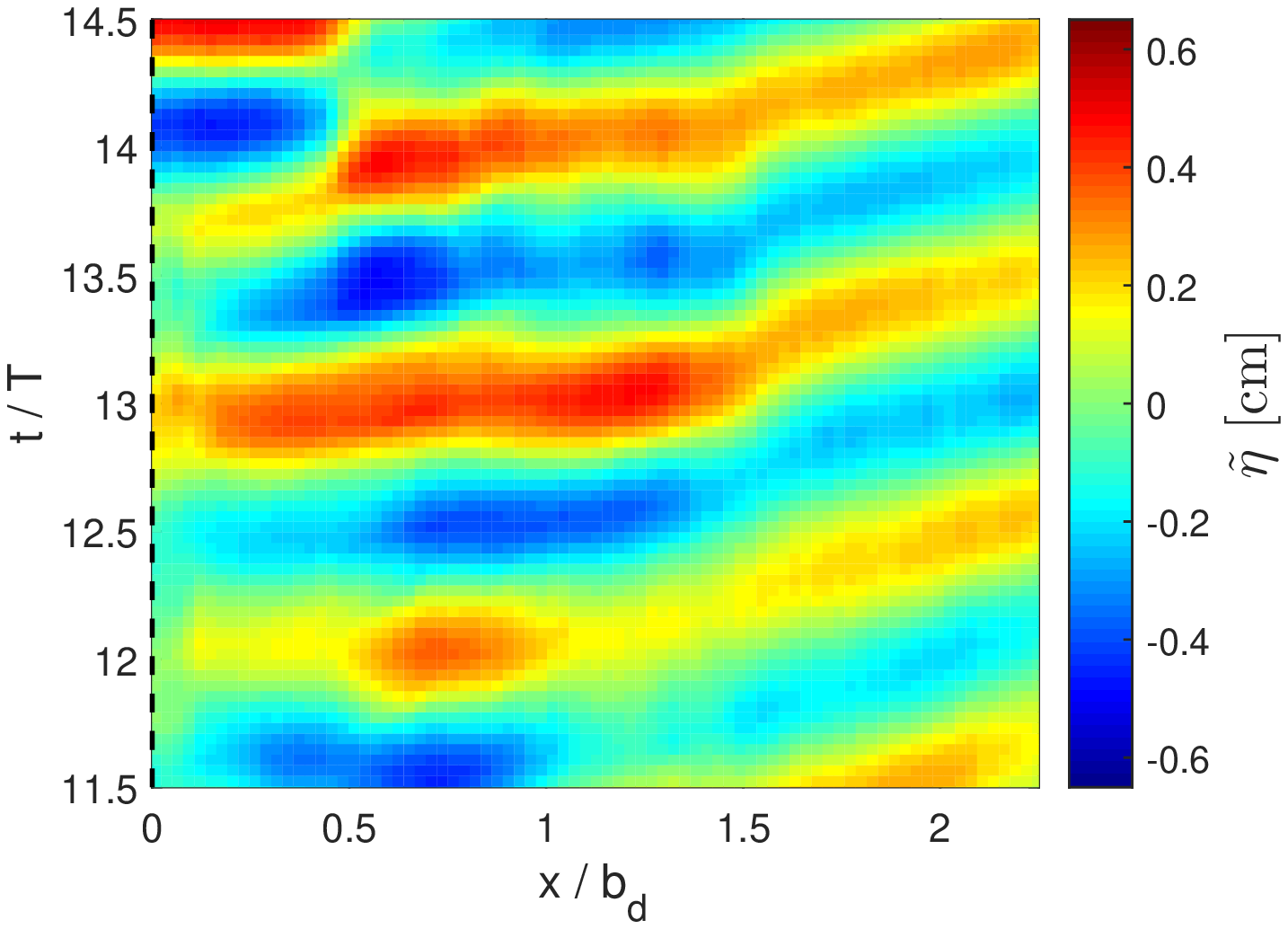}    } 
	    	
	    \caption{(a)  Spatiotemporal diagram of the height perturbation  $\eta(x,t)=h(x,t)-\langle h \rangle$, filtered in the frequency range $f \in [0.3-0.4]$Hz. The abscissae are rescaled by the radius of the dome $b_d$, the time by the mean wave period $T=2.9s$ and, the amplitude of the height   by the depth of the dome $\xi$. (b) Zoom on the region $x \in[0-2.25]  b_d$ and $t \in [11.5-14.5]T$.    \label{fig:interface_ampl2}}
	 \end{figure}

 We also    calculate    the root mean square amplitude of the wave
 \begin{equation}
A_\eta(x)= \sqrt{2} \left[ \frac{1}{T_e} {\int_0}^{T_e} \eta^2 (x,t)  t \right]^{1/2}
\label{def:ampl_a}
\end{equation}	 
	 
\noindent with  the height perturbation  $\eta$ without the slow modes component suppressed by a high-pass filter and $T_e$ the duration of the measurement.  We report $A_\eta$ (upper dotted   curve, called total) as a function of $x/b_d$ on the   figure \ref{fig:onde_vitesse_ampl}(b). The    dashed curve (called fundamental) corresponds to the amplitude of  the waves   in the frequency range $[0.3-0.4]$Hz (by using a band-pass filter) and the heavy curve corresponds to the amplitude with higher frequencies. {   Despite the fact that the typical length scale of the waves is comparable with the  distance of propagation, we distinguish different  behaviors of the amplitude of the wave during the propagation. } We have reported the three regions of interest: the region I, where the waves are generated and amplified; the region II, where the waves break, and the region III out of the dome.   The amplitude of the fundamental frequency increases in the region I ($\vert x \vert <0.565 b_d$), suddenly drops in the region II, and then decreases monotonically out of the dome  $\vert x \vert> b_d$. In the breaking  region, the amplitude of the modes with  large frequencies decreases just after those of the fundamentals. The shift of the amplitude drop is a consequence of the wave breaking, where the density patches are  wrapped and  thinned (figure \ref{fig:image_erosion1}(b)).

	  \begin{figure}   
	  		\hbox{ \centerline{(a) \hspace{6cm} (b)}}
	  			     \includegraphics[width=6.75cm,height=5cm]{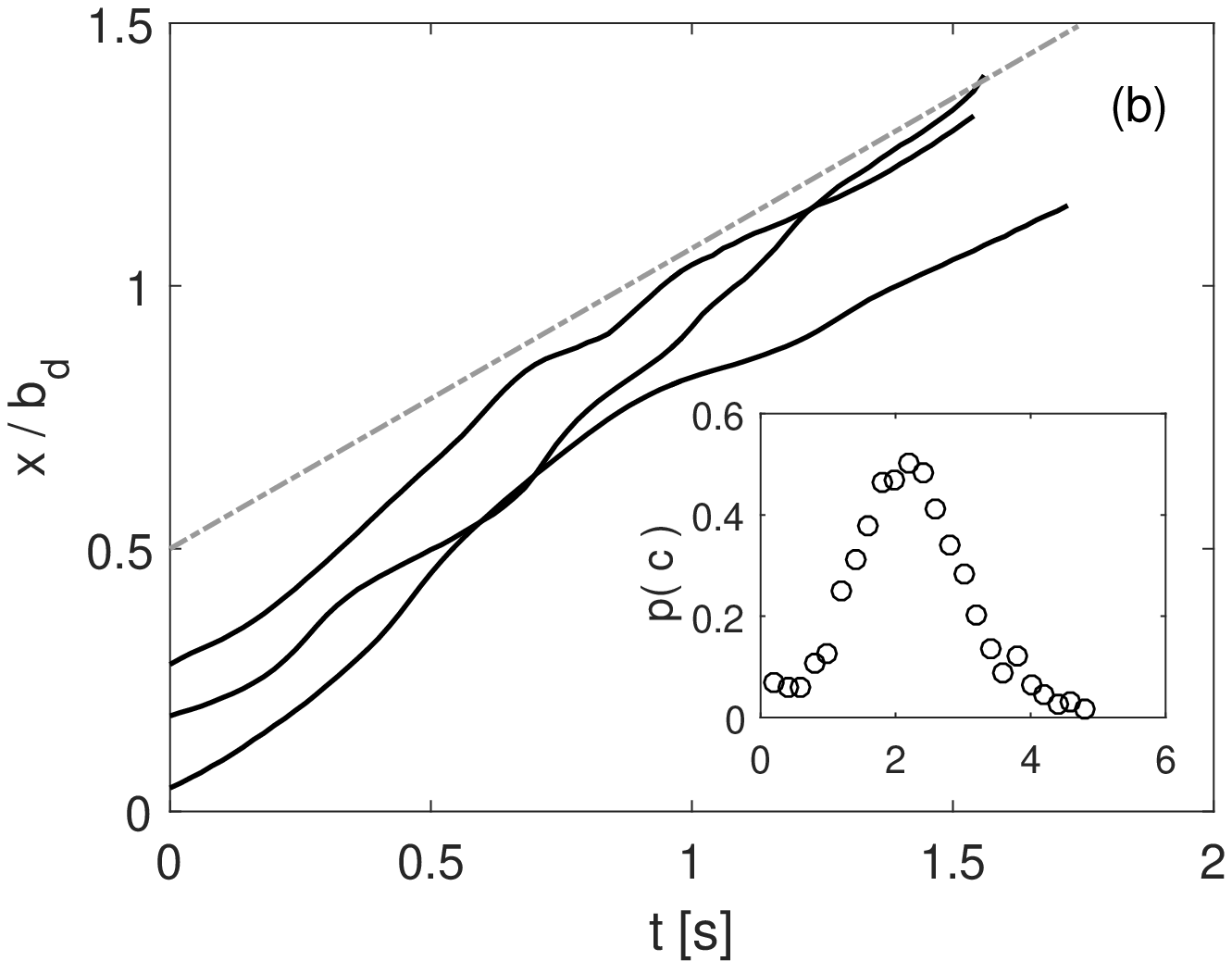} 
	           \includegraphics[width=6.75cm,height=5cm]{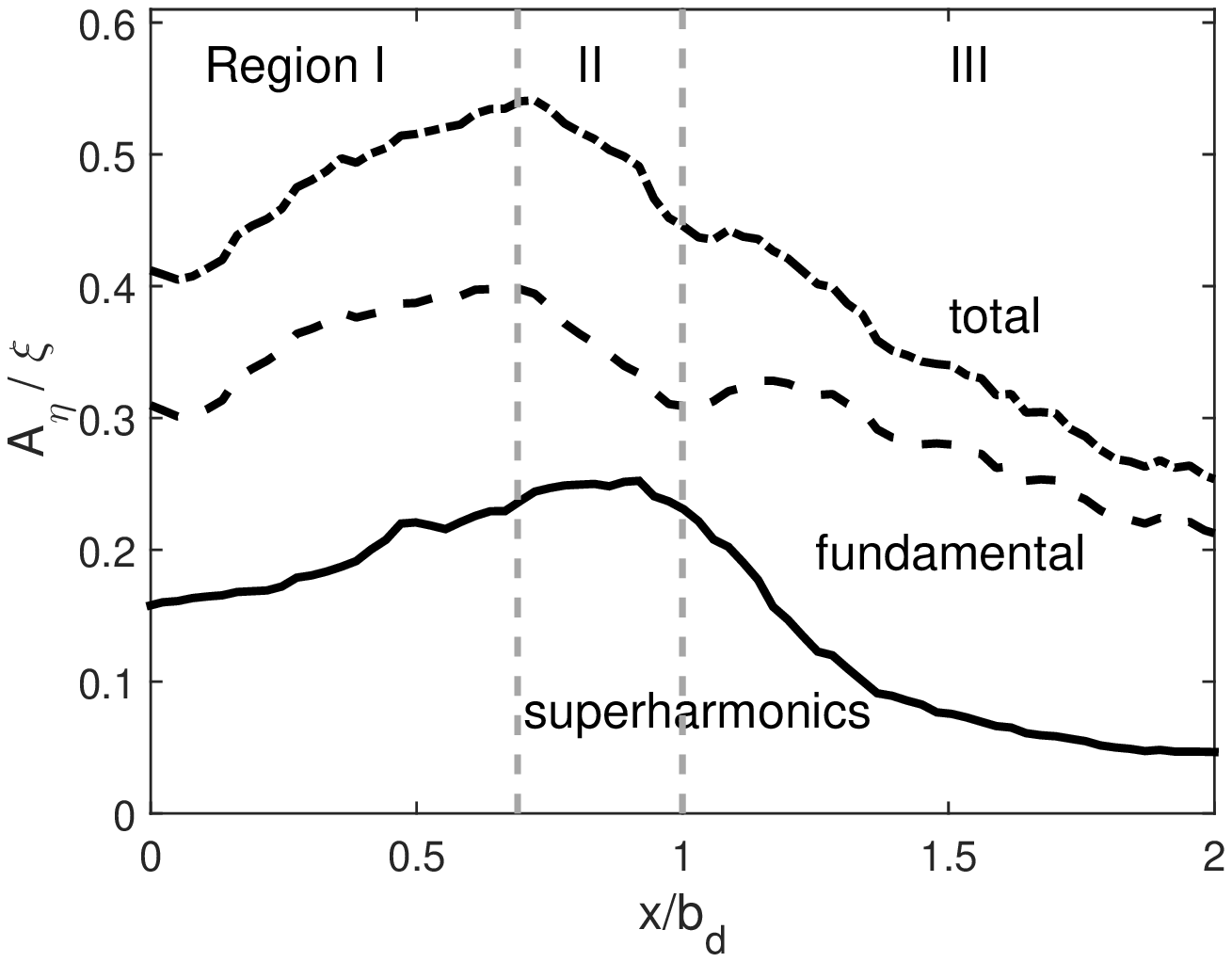}

	    \caption{(a)  Three trajectories of the wave envelop  (heavy lines), compared to a virtual trajectory (dashed line) with  constant phase velocity $c_p=2$cm/s. In the insert: distribution of the measured instantaneous phase velocity $c$.  (b) Amplitude of the wave $A_\eta$  given by   (\ref{def:ampl_a}). From top to bottom, the three curves show the total amplitude of the waves, the fundamental component and the higher harmonics.   \label{fig:onde_vitesse_ampl}}   
	 \end{figure} 

	\subsection{Wave generation}
\label{sec:generation}

	We now focus on the waves excitation mechanism, with two possible candidates: the instability of the interface   and the interface  excitation by turbulent fluctuations. First, we investigate the stability of the interface and we demonstrate that the shear at the interface is not strong enough  to trigger   Kelvin-Helmholtz and Holmboe instabilities. 
	
	{ 	The stability of the interface is related to the local Richardson number given by $Ri=  N^2/s_0^2  $ with $s_0$  the shear defined by $s_0=(\partial_{z_1} U_t)$ at $z_1=0$ and $N$ the frequency of interfacial gravity waves.  It compares the destabilizing effect of the shear to the stabilizing buoyant effects.  The perturbations being  interfacial  gravity waves, the Richardson number is defined by  $Ri=  (g '/  b_d)/s_0^2  $ \citep{chandrasekhar1961hydrodynamic}. It is worth noting that by using the  frequency of   internal gravity waves instead of the one of   interfacial  waves,  the Richardson number becomes one order of magnitude larger that the one defined here.

}
  
	 {  The shear $s_0$ is calculated from the mean velocity field and  corresponds to the tangential   velocity  gradient at the interface $\langle h \rangle$ (figure \ref{fig:velocity_jet11}(b)). We have shown that the interface merges with a streamline of the average flow passing close to the stagnation point (figure \ref{fig:velocity_jet1}). The mean velocity field along the interface satisfies the relation  $\langle u_x \rangle \partial_x \langle h\rangle \simeq \langle u_z \rangle$ for $\vert x \vert<0.6 b_d$, which  implies that $\partial_t \langle h\rangle \simeq 0$ by considering the continuity of the mean material line.  The average velocity field and the mean position of the interface may be considered in   quasi-equilibrium, supporting the stability analysis of the interface from the mean fields.}

	 The local Richardson number is reported on  the   figure \ref{fig:velocity_jet2}(b). It varies between  $40$ and $230$  in the vicinity of the head of the dome, i.e. $|x|<2.5cm$  or $0.61 b_d$. It turns out that the interface should not be destabilized in this region by a classical Kelvin-Helmholtz instability   operating for $Ri <1/4$ \citep{miles1961stability,chandrasekhar1961hydrodynamic}(see also the numerical simulations by \cite{woodward2014hydrodynamic} on a closely related problem).  We point out that the presence of a curved interface and an accelerating flow, i.e $\partial_x U_t>0$, may change the criteria of stability. Nonetheless these effects are only significant on the sides of the dome, where the waves are already excited and amplified.

	 {   Moreover, the length scale of the instability is not compatible with the one of the Kelvin-Helmholtz instability. For a sharp velocity jump, the  smallest unstable wave number is given by \citep{chandrasekhar1961hydrodynamic}
	 
	 \begin{equation}
	 k_{min} \simeq \frac{2 g At}{\Delta U^2}
	 \end{equation}
	 
	 \noindent with $\Delta U$ the velocity jump at the interface and $At=(\rho_2-\rho_1)/(\rho_2+\rho_1)$ the Atwood number. From figure \ref{fig:velocity_jet11}(b), the largest velocity jump is typically $\Delta U_{max}=1 $cm/s inside the dome. The numerical application gives $k_{min}=5.8  \times 10^{3}$ m ${}^{-1}$ corresponding to a wavelength smaller than $0.1$ cm. The Kelvin-Helmholtz instability could only trigger perturbations at scales  $20-40$ times smaller than the one observed.  Reciprocally, the velocity jump  should be at least $6$ times larger to account for the observed wavelength. }	{   During the erosion process, $Ri$   decreases \citep{kumagai1984turbulent}: if the interface is initially stable for the Kelvin-Helmholtz instabilities, it remains so during the erosion process.}

	 
	 An other alternative to  the Kelvin-Helmholtz instability is the Holmboe instability \citep{holmboe1962behavior}. This instability can operate for large  Richardson numbers and \cite{alexakis2009stratified} showed that it may appear for $Ri=30$ (far from the $1/4$ of the K-H instability). However, the Richardson number is larger than $40$ for $\vert x/b_d \vert<0.6$ in the dome. {   The mode associated with the Holmboe instability is also characterized by symmetric cusps   at large Richardson number \citep{strang2001entrainment}. This pattern   has not  been observed in  our experiment.} We conclude that neither the Kelvin-Helmholtz instability nor the  Holmboe instability   generate perturbations on the interface.



 	 \begin{figure}  
 	\hbox{ \centerline{(a) \hspace{6cm} (b)}}
 	     \includegraphics[width=6.75cm,height=5cm]{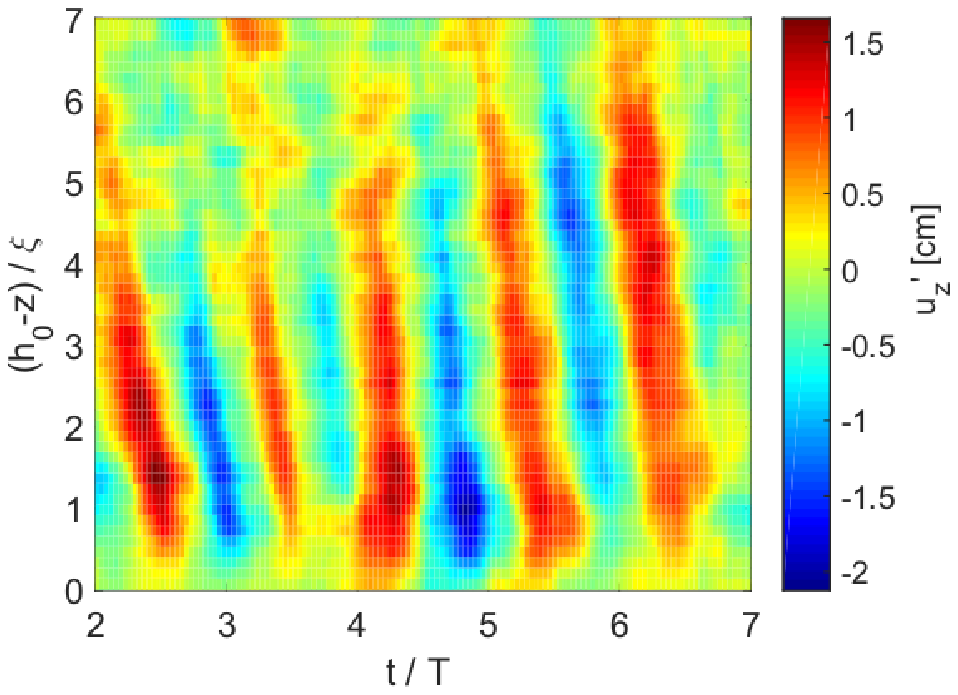} 
 	              \includegraphics[width=6.75cm,height=5cm]{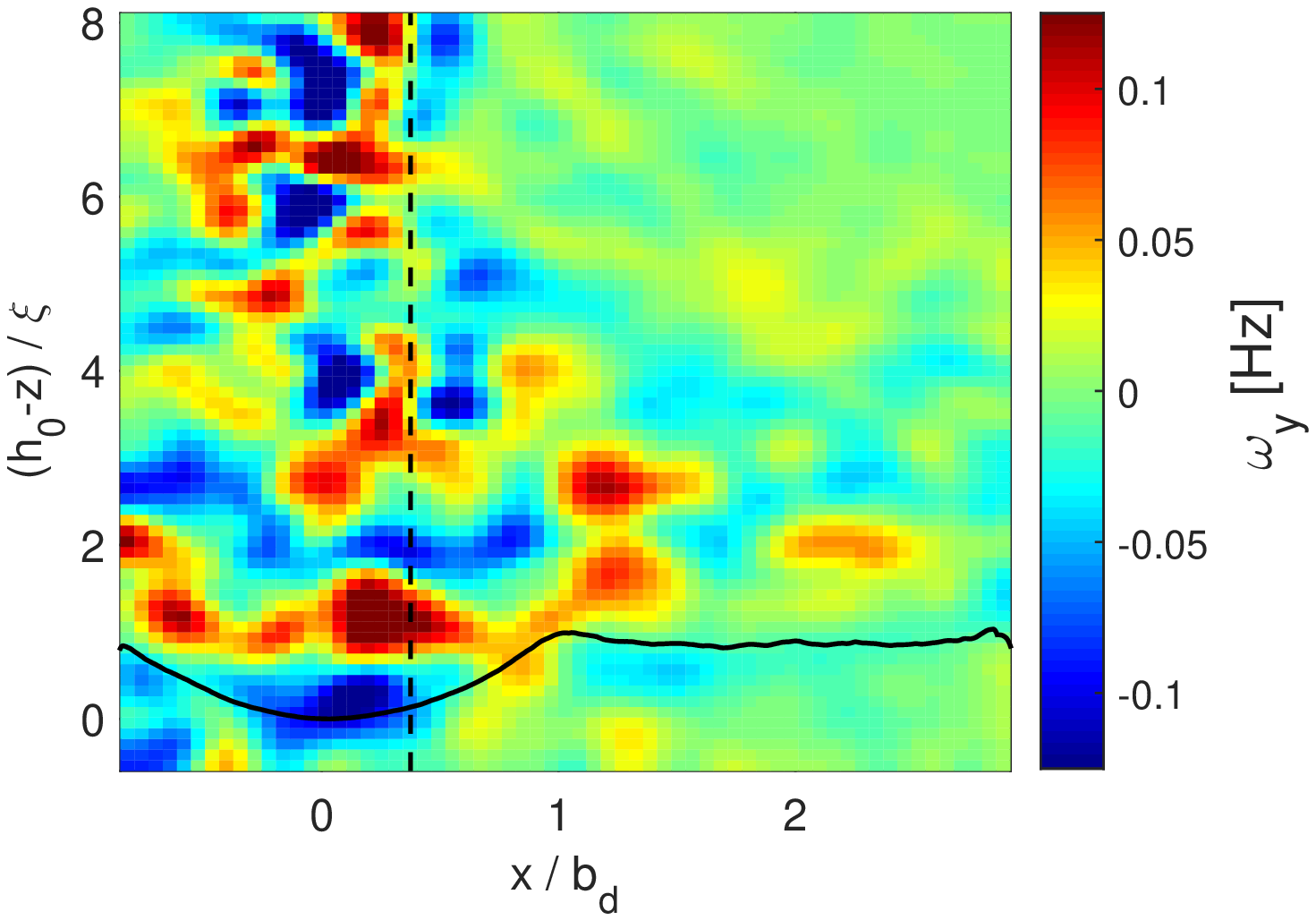} 
 	    \caption{(a) Spatiotemporal diagram of the fluctuations $u_z'$  in the range of frequencies $[0.3,0.4]$Hz  along the vertical axis $x=0.37 b_d$. (b) Bandpass filtered fluctuations of the vorticity field, calculated from a coarse-grained  velocity field  with width of $0.75cm$ (Gaussian filter). The dashed curve represents the vertical axis at  $x=0.37 b_d$. \label{fig:uz_omega}  }  
 	 \end{figure}
	 
	The source of waves is  ensured by the turbulent fluctuations of the jet. Having identified the region and the frequency range of the waves in   section \ref{sec:properties}, we look for their signature in the turbulent jet.   In order to demonstrate that the perturbations come  from the jet, we focus on the  fluctuations of {   the vertical  velocity fluctuation} $u_z'=u_z-\langle u_z \rangle$ along the axis $x=0.37 b_d$, represented by the dashed curve on    figure \ref{fig:uz_omega}(b). We calculate the bandpass filtered component of $u_z'$ in  the frequency range  $[0.3-0.4]$Hz. The associated  spatio-temporal diagram is reported on the   figure  \ref{fig:uz_omega}(a) with $t$ and $z$ rescaled by the period of the wave $T$ and the depth of the dome $\xi$. We clearly identify perturbations propagating from the jet to  the dome. It is worth noting that these fluctuations appear far from the dome at $\vert h_0- z\vert \simeq (5\pm 1) \xi$ and   seem  to be incoherent before. This region is far from the transitional distance  close to the nozzle of the jet ($z<50 b_0$), where coherent structures may be generated in the early development of the jet \citep{yule1978large}.  Thus, it is possible that a coherent process, based on the interplay between the jet and the interface, may filter or select the  perturbations far from the interface. The  bandpass filtered fluctuation of vorticity field, calculated from a coarse-grained  velocity field, is reported on   figure \ref{fig:uz_omega}(b) for a given time $t$. The fluctuations $u_z'$ are associated with large scale vortices, which propagate toward the interface. The diameter of the vortices close to the dome is varying around $\lambda= 3 \pm 1$cm. At the entrance of the dome, the typical vertical velocity is $\langle u_z \rangle \simeq 2 \pm 1 cm/s$ (figure \ref{fig:velocity_jet1}). So a typical period of excitation corresponds to the advection of two vortices of opposite sign and  the associated time scale of forcing is $2 \lambda/\langle u_z \rangle \simeq 3$s, a value very close to the wave-period $T$. 
	  
 	  From the fluctuation $u_z'$, we can estimate a Richardson number for the observed eddies. The maximum  amplitude of  $u_z'$ reaches typical value given by $u_ {max}= 2$cm/s. The Richardson number of these eddies is defined by $Ri_\lambda=g' \lambda /(u_ {max}^2)$ \citep{cotel1996model} and corresponds to   $Ri_\lambda \simeq 22$. An energy balance between the kinetic energy of the vortex and the potential energy associated with the penetration depth $\delta$ gives $\delta/ \lambda \sim Ri_{\lambda}^{-1}$ \citep{linden1973interaction}. From the measured Richardson number, the penetration depth is of order $\delta / \lambda \simeq 10^{-2}$. Thus the convolution of the interface is expected to be too small if we consider the classical model of the vortex impact. It has been recently demonstrated in the context of internal-gravity waves excited by turbulent plume that   waves can be forced by the bulk pressure-field of the turbulence \citep{lecoanet2015numerical}. This approach could be fruitful to explain the wave excitation for eddies with low Richardson numbers.


	\subsection{Wave amplification}
	\label{sec:amplification}
	 
The observed growth of the  perturbations along the interface (figure \ref{fig:onde_vitesse_ampl}(b)) may be explained  by a mechanism of wave amplification. Furthermore,  the axisymmetry of the configuration  leads a priori to a decrease of the energy of the perturbations, due to the radial spreading of the energy. Thus, the amplification must at least counterbalance the energy spreading due to the axisymmetry. As shown in the previous section, the amplification mechanism could not be explained by the Kelvin-Helmholtz and Holmboe instabilities. It is interesting to  note that our configuration   shows some similarities with the problem of the generation of ocean-waves by turbulent wind \citep{janssen2004interaction}. The ocean-waves are generated by  turbulent pressure fluctuations with  a sheared mean flow. Like in our configuration, the Kelvin-Helmholtz is not the best candidate to explain the generation of ocean-waves \citep{janssen2004interaction}. Focussing initially on an ideal configuration of a shear flow with a density jump \citep{miles1957generation},  \cite{miles1960generation} showed that  the presence of a shear flow may lead to an enhancement of the energy transfer to the waves in turbulent shear flows. This scenario is based on the presence of a critical layer \citep{miles1957generation}  at the height $z_c$ where the phase velocity of the waves matches the velocity $U_t(z_c)$.  It is usually called the Miles instability.  The Miles instability permits the transfer of energy from the mean flow to  the wave  if at the critical height $z_c$, the second derivative of $U_t$ is negative. {  This instability is characterized by waves induced stress, where the Reynolds stresses $\tau=-\langle u_x' u_x'\rangle$ of the velocity fields  are non-zero}. A physical interpretation of the instability  is given by \cite{lighthill1962physical}. This mechanism is similar to the critical layer encountered in shear flow without inflection point \citep{drazin2004hydrodynamic}, but it differs from the critical layer found in stratified flows, where the internal gravity wave gives its energy to the mean flow, even if the amplitude of the wave may increase due to the wave-action conservation \citep{sutherland2010internal}. {  Our measurements show that the mechanism of amplification operating in the dome shares some features with  the Miles instability. Despite the difference in geometry between the framework of the  Miles instability and our configuration, we use the theoretical result of the Miles instability, in the absence of theoretical prediction adapted to our case.} 	
	

	 \begin{figure}  
	 	\hbox{ \centerline{(a) \hspace{6cm} (b)}}
	       \includegraphics[width=6.75cm,height=5cm]{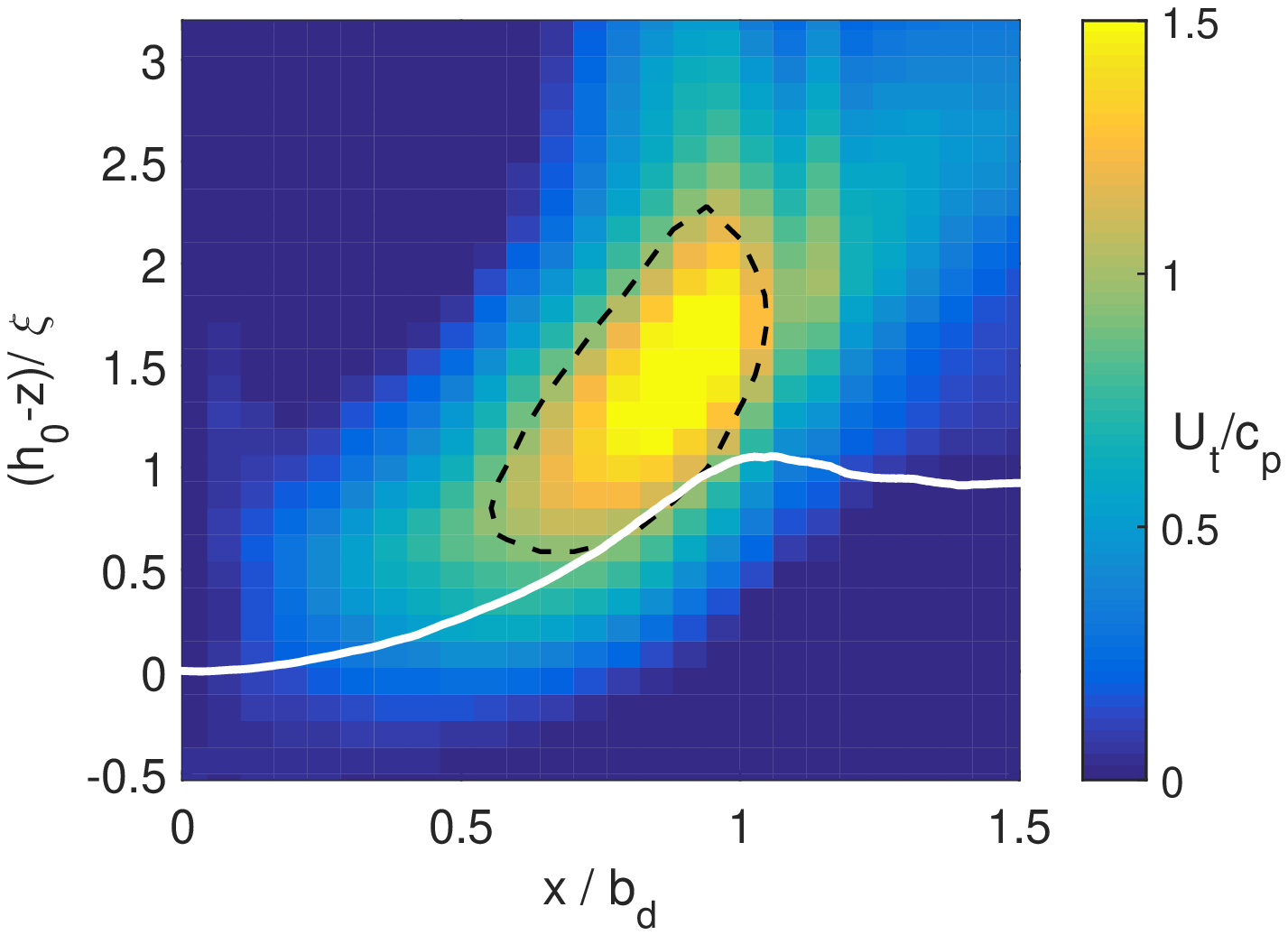}  
        \includegraphics[width=6.75cm,height=5cm]{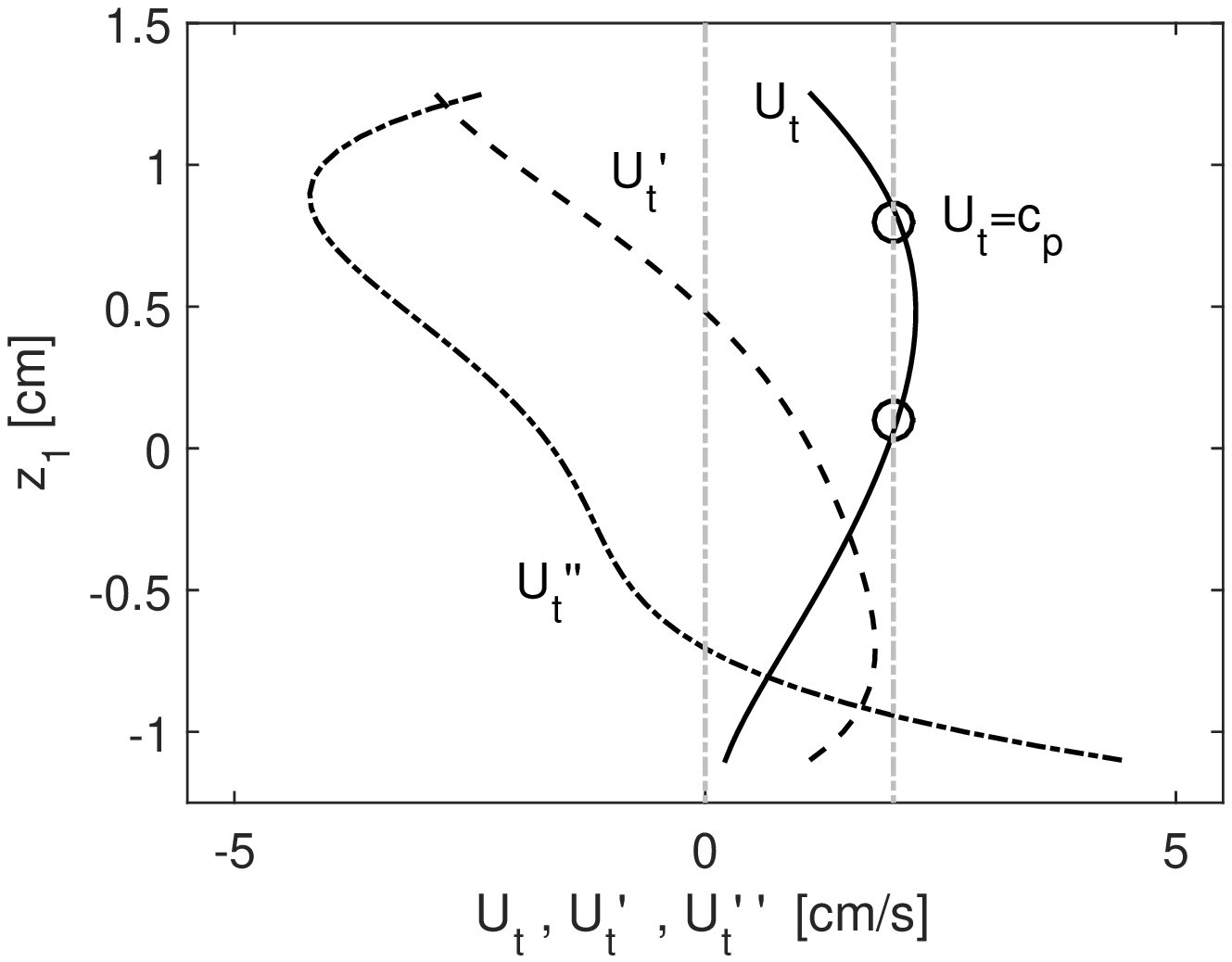}   
	   \caption{(a)  Tangential velocity $U_t$ divided by the phase velocity $c_p=2cm/s$. The black dashed curve represents the contour where  $U_t=2cm/s$. The white curve corresponds to the interface. (b) Profile of  the tangential velocity $U_t$ corresponding to the abscissa $x=0.7  b_d$ extracted from  figure \ref{fig:velocity_jet11}(b). Its first $U_t'$ and second   $U_t''$ depth derivatives  are represented respectively by   dashed and dotted lines. The circles correspond  to the points where the velocity is equal to $c_p$. \label{fig:couche_critique}}  
 
	 \end{figure}

\begin{figure}  
 	     \centerline{ \includegraphics[width=11cm,height=6cm]{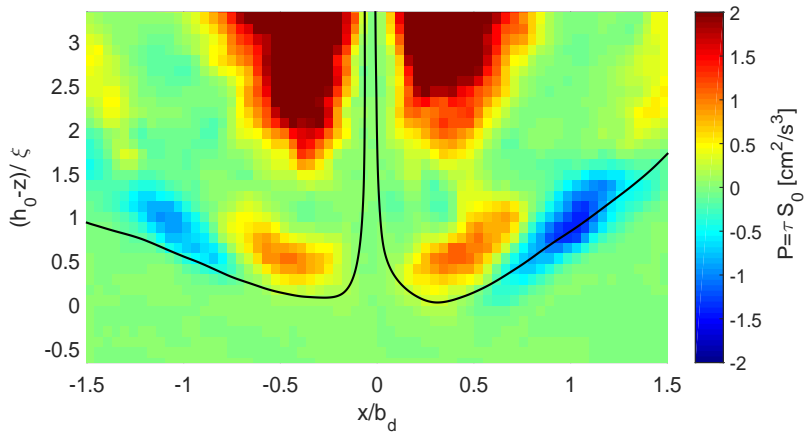}  }
	   \caption{Production term $P(x,z)=\tau  S_0$ with $\tau= - \langle u_x' u_z'\rangle$ the Reynolds stress and $S_0$ the large scale shear. The   black curve corresponds to the streamline passing close to the stagnation point.   \label{fig:couche_critique2}}  
 
	 \end{figure}

\begin{figure}    
 	     \centerline{ 
 	      \includegraphics[width=8cm,height=5cm]{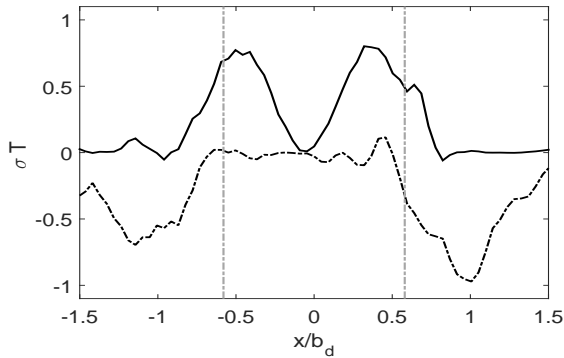}     }
 	     
	   \caption{Maxima (heavy line) and   minima (dashed line) of the dimensionless energy transfer rate $\sigma  =  P/(2 E)$ rescaled by the wave-period $T$, where $P$ is the production term and $E$ the kinetic energy,   for all heights $z$ in the dome as a function of $x/b_d$. The grey lines correspond to the abscissa $x \pm 0.58 b_d$  where the waves may propagate slower than the mean flow.    \label{fig:sigma}}  
 
	 \end{figure}

First, we look for the region of resonance where the typical phase velocity $c_p$ and the tangential velocity $U_t$ are equal.  In figure \ref{fig:couche_critique}(a), the velocity  $U_t$ is divided by the    phase velocity $c_p=2cm/s$ for the half plane of  symmetry  of the jet given by  $x>0$. The  white curve is the mean interface $\langle h \rangle$. The closed dashed curve represents the contour  where  $U_t=c_p$: inside this region, the wave may propagate slower than the flow.  The enclosed region   appears for  $\vert x \vert>0.58 b_d$ and ceases for $x \simeq b_d$. If the Miles instability is present, an  increase of the  disturbance energy  must be seen through the action of the Reynolds stress \citep{lin1954some,drazin2004hydrodynamic}. The energy transfer between the mean flow and the waves is given by the production term $P(x,z)=\tau S_0$, with the large scale shear $S_0= \partial_x \langle u_z \rangle+\partial_z \langle u_x \rangle$ and the Reynolds stress $\tau=- \langle u_x' u_z' \rangle$.  The Miles instability must be associated with a positive production term in the region where $c_p>U_t$. The production term $P$ is calculated for the run M$3$ (see table \ref{tab:runs}) and it is reported on    figure \ref{fig:couche_critique2}. We observe that the production term is positive  for $\vert x \vert$ larger than $0.1 b_d$ and it becomes negative for $\vert x \vert$ in between $0.55$ $b_d$ and $0.75 b_d$, depending on the height.  The  transition between positive and negative production occurs in the region where $U_t \simeq c_p$  for  $\vert x \vert=0.58 b_d$  (dashed curve figure \ref{fig:couche_critique}(a)), suggesting the presence of a critical layer.  The production term is non-zero above the interface, as expected from  the Miles instability. The sign of the shear $S_0$ remaining constant for each side  of the dome ($S_0<0$ for $ x  <0$ and $S_0>0$ for $x>0$), the variation of $P$ is associated with a change of sign of the Reynolds stress $\tau$. It is a classical property of the critical layer \citep{miles1957generation,drazin2004hydrodynamic}.  The relative phase between $u_x'$ and $u_z'$ varies abruptly across the critical layer where $U_t=c_p$, leading to a change of the sign of the production term \citep{lin1954some}. The sign of $P$ is also  correlated to the variation of the height  $A_\eta$ of the waves (figure \ref{fig:onde_vitesse_ampl}(a)). The height of the wave increases when $P$ is positive and it  decreases when $P$ becomes negative.  
 From the production term $P$ and the kinetic energy of the wave $e=\langle u_x'^2+u_z'^2 \rangle/2$, an estimation of the energy transfer rate $\sigma$  of the wave is possible with  $\sigma=P/(2 e)$. We report the maxima (heavy black line) and the minima (dashed black line) of the dimensionless energy transfer rate $\sigma T$ for all heights $z$ in the dome as a function of $x/b_d$ on   figure \ref{fig:sigma}.   The positive energy transfer rate reaches the values $0.8 T^{-1}$ at $x=0.4 b_d$ and the damping rate reaches the values $-0.95 T^{-1}$ for $x=b_d$.  We clearly observe that the waves are amplified, with $\sigma>0$,  in the central region  and they are significantly damped outside, with $\sigma<0$ The grey lines correspond to the abscissa $x= \pm 0.58 b_d$  where the waves may propagate slower than the mean flow. Once again, the transition between amplification process and damping process  occurs in the region where the waves cross the critical layer.
 
%
%
 
  The calculation of the growth rate of the Miles instability is complex and an explicit formula exists only for idealized  configurations \citep{miles1957generation,morland1993effect,drazin2004hydrodynamic}. The growth rate depends on  properties of the mean flow at the critical layer and its calculation requires the knowledge of the amplitude of the initial neutral perturbation. It is out of reach of the present study and we will only focus on the sign of the growth rate in order to confirm the presence of the instability.  The grow rate   is given at leading order in $ \rho_1/\rho_2$ \citep{miles1957generation,drazin2004hydrodynamic,young2014generation}

\begin{equation}
 \sigma= -\frac{\pi}{2} \frac{\rho_1}{\rho_2}    c_p   \left(\frac{d^2 U_t}{d {z_1}^2} \right) \left(   \frac{d U_t}{d z_1}  \right)^{-1}  \left \vert \frac{\phi_c}{\phi_s} \right \vert^2
\label{def:sign_sigma}
\end{equation}

\noindent with  $\vert \phi_c/ \phi_s\vert$ the ratio of the streamfunction at  the critical layer, which is especially difficult to estimate. Considering wave-amplification, the growth rate may be interpreted as a   energy transfer rate from the mean flow to the waves  \citep{miles1960generation}.  The   criteria of instability requires that the second derivative of $U_t$ must be negative. In figure  \ref{fig:velocity_jet11}(b), the profiles of $U_t$ are concave, which suggests that $U_t''=(d^2 U_t)/(d {z_1}^2)$ is negative. We verify this properties for the profile at $x=0.7 b_d$ (fifth profile starting from the left of   figure  \ref{fig:velocity_jet11}), where   $U_t=c_p$ for $z_1$ equal to $0.1$ and $0.85$cm. The first derivative $U_t'$ and the second derivative $U_t''$ are reported on  figure \ref{fig:couche_critique}(b). The derivative $U_t''$  remains negative in the region where the wave can be in resonance with the mean flow. The derivative $U_t'$ is positive for $z_1<0.5$cm.  The sign of the ratio $U_t''/U_t'$ being negative, the mechanism of the Miles instability may    amplify the wave.   {  Our result shows that the amplification mechanism shares   features with   the Miles instability, suggesting the presence of a critical layer, that amplifies the waves generated by the eddies from the jet}. 

	We have shown in the previous section that the waves are generated by the eddies of the jet and not by an instability. It could seem contradictory to the present section, which aims to demonstrate the presence of an instability. However, both mechanisms are complementary: the turbulence excites the waves and the instability amplifies them. It is a  scenario similar to the  ocean-wave generation mechanism. The opposite scenario may be addressed, i.e the waves are triggered by the instability and amplified by the turbulence. However, it is not very likely that the Miles instability may trigger a wave on a distance comparable to the half-wavelength. For the last point,  \cite{phillips1957generation} has considered the wave-amplification by a resonance between the turbulent pressure field and the waves. {  By taking into account the presence of a critical layer, \cite{miles1960generation} has    shown that this process is strongly enhanced by the Miles instability.} Thus, once the wave is triggered, the energy transfer from the mean flow should be significantly larger than the energy coming from the turbulent fluctuations.

	\subsection{Wave breaking}
	\label{sec:breaking}
	
	We now focus on the wave breaking occurring in the region $0.6< \vert x/b_d \vert<1$. In this area,  different elements may explain the wave breaking. First, we have shown that when the phase velocity becomes larger than the mean velocity, the negative Reynolds stress damps the wave due to the critical layer. Then, the profile of $U_t$ becomes much steeper for $ \vert x \vert >0.6 b_d $  (figure \ref{fig:velocity_jet11}(b)). The associated local Richardson numbers (insert in figure \ref{fig:velocity_jet2}(b))  are  of order unity in this region, which may lead to a destabilization of the wave by  a shear-instability. Finally, the steepness of the wave, which is defined by the product of the wavenumber $k_0$ with the height $A_\eta$, becomes close to the critical steepness of the wave  \citep{babanin2011breaking}. We consider the maximum average height of the wave  (dotted curve indexed by \textit{total}, figure \ref{fig:onde_vitesse_ampl}(a)), which reaches the values $  A_\eta ^{max}  \simeq 0.56 \xi$ with $\xi \simeq 0.8$cm, i.e  $  A_\eta ^{max}=0.45$ cm,  for the run M$2$ (table \ref{tab:runs}).  {   The maximum   steepness based on the smallest wavenumber $k=0.5 k_0$, where $k_0=2 \pi/b_d$ (run M2, see tab.\ref{tab:runs}), is $k A_\eta^{max} \simeq  0.35$, a value in the range of wave-breaking criteria   $k A_\eta^{max}  \in [0.3-0.4]$ for ocean-waves \citep{babanin2011breaking}}. The damping, the shear and the large steepness of the wave may lead to the wave breaking. It is difficult to disentangle the different effects, which can of course interplay between each others \citep{banner2002determining}.

	 After the breaking region just  above the interface (figure \ref{fig:interface_ampl}(a) for $x \in[4,6]$ cm and $h_0-z \in[1,3]$ cm), we observe an area of mixed fluid with a relative density $\hat \rho$ varying between  $0.4$ and $0.2$. It shows that the breaking has already induced  mixing. Indeed, wave-breaking is believed to be the main mechanism of mixing at low Froude number \citep{fernando1991turbulent,fernando1996some}. The  mixed fluid is then transported upstream, following the streamlines of the mean flow (figure \ref{fig:velocity_jet1}). Figure \ref{fig:image_erosion1}(a) shows that the mixing continues at the frontier of the jet, where the turbulent intensity is expected to be important \citep{list1982turbulent}. Due to the lateral entrainment of the jet (section \ref{sec:jetstratif}), a part of the mixed fluid is directly injected inside the turbulent jet.

	
	\subsection{Remark on the length and time scales}
	
{ 	We have studied the different properties  of the generation, the amplification and the breaking of the waves in the previous sections. These different mechanisms take place on a distance, i.e. the width of the dome,  comparable to the wavelength of the wave. In the framework of the Miles instability, the long-wavelength perturbations are the most unstable \citep{miles1957generation,morland1993effect,drazin2004hydrodynamic}. This mechanism could explain the observed large wavelengths in our confined geometry. From the lagragian point of view, the waves are generated and destroyed on time scales comparable to their period of oscillation.  Figure \ref{fig:sigma}  confirms that the    evaluated energy transfer  rate between the waves and the mean flow  is close to the frequency of the wave. Unlike classical studies of waves propagation, the times and length scales are not well decoupled, which is a limitation of the local study of the stability of the interface. 
	
	Our study shows  that  the  wavelength, the frequency and the growth-rate of the waves are determined by    the forcing   and the amplification mechanisms. In   section \ref{sec:amplification}, we have shown that both mechanisms are not independent: the perturbations propagating  toward the dome oscillate at the wave frequency.  A possible explanation  is the existence of a global mode, coupling the dynamics of  the interface and the  flow outside the dome. The limitation of our local study may be due to the interpretation of a global mode in term of local perturbations. More   work is needed for the theoretical analysis of the observed results, but we think the basic physical processes are correctly described here. }
 
\section{A scaling  law for the entrainment rate}
\label{sec:model_law}
 
 \subsection{The scaling  law  }
 
\begin{figure} 
\hbox{ \centerline{(a) \hspace{6cm} (b)}} 
\centerline{ \includegraphics[width=6cm,height=4cm]{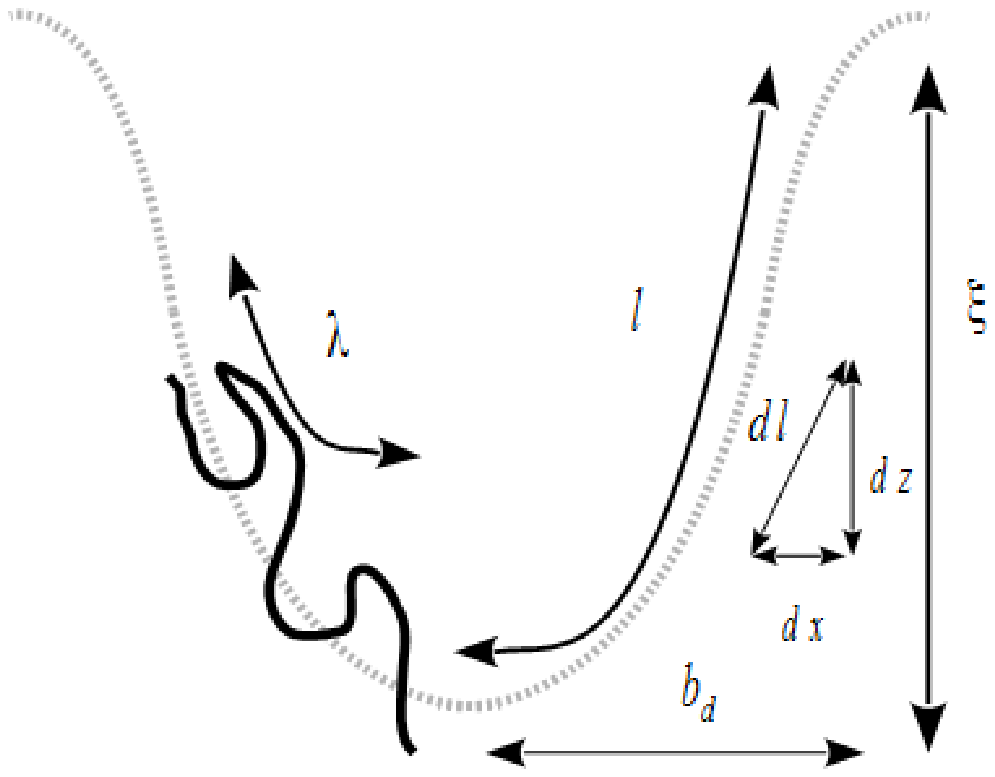}
\includegraphics[width=7cm,height=4cm]{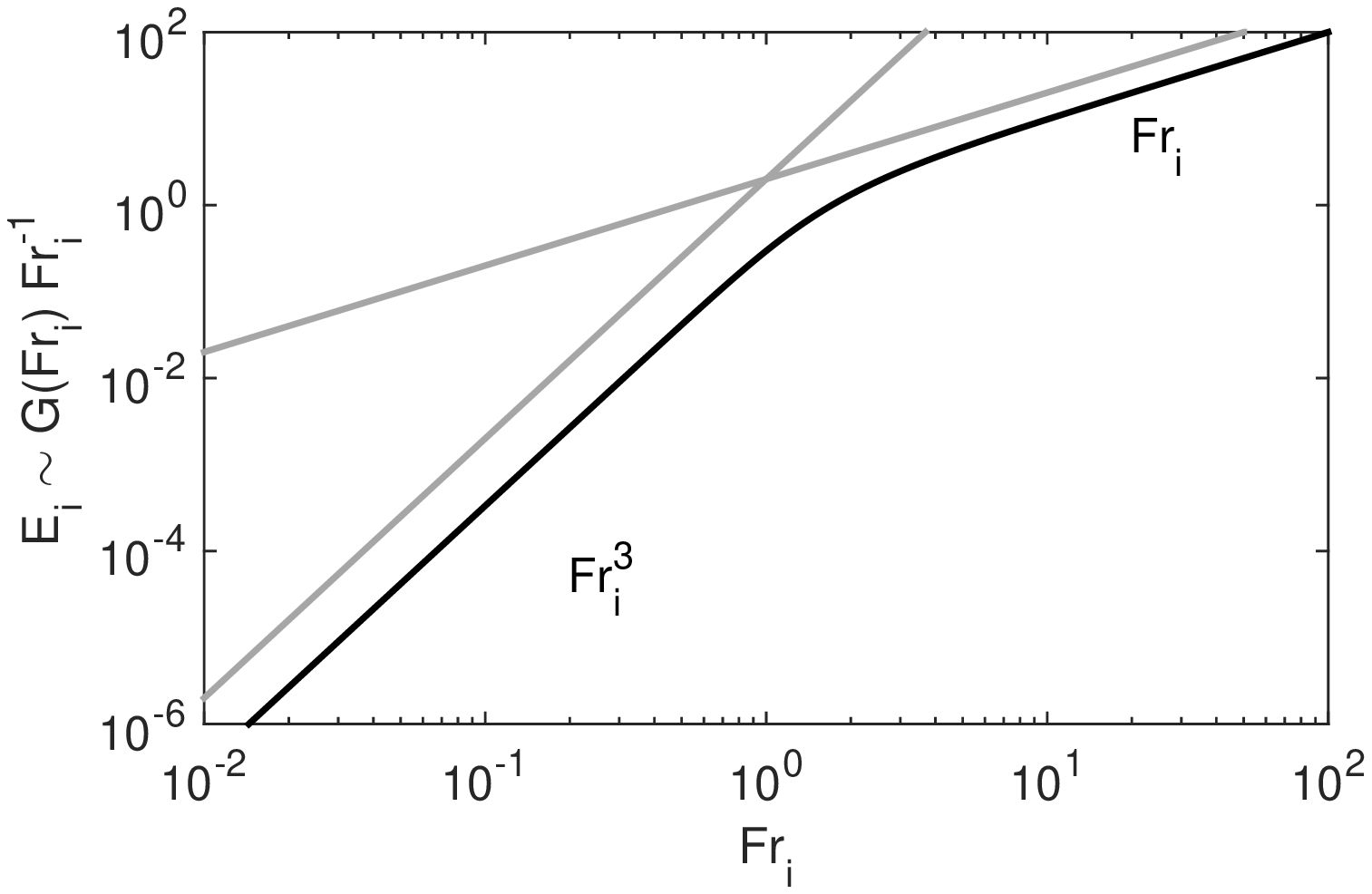}  }
\caption{(a) Schematic and notation  for the model of entrainement rate. (b) Entrainement rate as a function of the Froude number given by   equation (\ref{def:eq_Ei}). The grey curves show the $Fr_i^3$ power law valid at low Froude number and the linear scaling $Fr_i$ valid at large Froude numbers. \label{fig:dome}}  
\end{figure}

{  In the previous sections, we have demonstrated that the erosion is related to the dynamics of gravity interfacial waves. We want to address in this section the scaling law for the erosion rate, given by equation \ref{eq:scalinglaw},  as a function of the Froude number  based on our observations.} \cite{linden1973interaction}, \cite{cotel1996model} and \cite{shrinivas2014unconfined} have  suggested different physical arguments for  the scaling law, leading to $E_i \sim Fr_i^3$ for the two first authors and $E_i \sim Fr_i^2$ for the later authors for $Fr_i<1$. These models have not considered  mechanisms involving waves for the erosion process,  except for a footnote (page  476) in  \cite{linden1973interaction} from Dr E. J. Hinch. Based on the data of    \cite{baines1975entrainment} and \cite{baines1993turbulent}, the entrainment rate should follow a $Fr_i^3$ scaling law at low Froude number and a linear scaling law at large Froude number. The volumetric flux entrained across the interface $Q_e$   (equation \ref{eq:scalinglaw}) varies with the vertical entrainment velocity $u_e$ defined as the amount of fluid entrained per unit height per unit time    \citep{linden1973interaction}. The entrainment   rate, defined in equation (\ref{eq:scalinglaw}), varies with the ratio $u_e/u_i$ \citep{cotel1996model}
 
 \begin{equation}
\frac{u_e}{u_i} \sim \frac{l_e f_e}{u_i}
 \label{def:entr_model}
 \end{equation}
	
\noindent where $l_e$ and $f_e$ are  typical length  and time associated with the erosion process and $u_e$ the vertical velocity of the interface. As we consider waves amplified by an unstable mechanism, $l_e$ is related to the relative length over which the amplification takes place,  and $f_e$  to the frequency of the wave-perturbation. We have shown that the waves are triggered by the eddies in the jet with time scale comparable to the buoyancy frequency, i.e. $f_e=\sqrt{g'/b_i}$. We choose the length $l_e$ as the distance $l$ travelled by the wave inside the dome minus the dome radius, i.e. $l_e=l-b_d$ (see fig.\ref{fig:dome}(a)).

 {  The waves exchange energy with the mean flow during the propagation along  the distance $l$ (section \ref{sec:amplification}). Hence, all the steps leading to the breaking of the waves take  place on this distance.} {  The length $l_e=l-b_d$  corresponds to a ``useful`` length in the sense that for $Fr_i \rightarrow 0$, no distortion and no erosion are expected in our model.  This mathematical regularisation prevents the divergence of $E_i$ for $Fr_i \rightarrow 0$,   a regime not considered here. \cite{cotel1997laboratory} report a  constant erosion rate $E_i$ for $Fr_i<0.14$, when the interface is almost flat. Our model quantifies the departure  from this regime. Thus, we construct the   length $l_e$, so that it corresponds to the additional length travelled by the wave,  when the dome is deepening for a constant dome radius $b_d$.}  The infinitesimal length $d l$ travelled by the wave on a interval $dx$ is $d l = \sqrt{dz(x) ^2+dx^2}$ (figure \ref{fig:dome}(a)) and the  useful length $l_e $ is
 
 \begin{equation}
 l_e = \int_{x=0} ^{b_d} \left[ \sqrt{ \left(\frac{d \langle h \rangle}{d x} \right)^2+1 }\right] d x-b_d
 \end{equation}

	\noindent with $\langle h \rangle(x)$ the depth of the dome as a function of $x$. Our measurements show  that $\langle h \rangle(x)=\xi (x/b_d)^2$, with $\xi$ the depth of the dome. By using the change of variable $x=u b_d$, the integral becomes
	
	 \begin{equation}
 l_e =b_d \int_0 ^{u=1} \left[ \sqrt{(\epsilon u)^2+1 }\right] d u-b_d
 \end{equation}
 
 \noindent with $\epsilon=\xi /(2b_d)$. The integration gives

 \begin{equation}
 l_e =\frac{b_d}{2} \left[ \sqrt{(\epsilon  )^2+1 }+\frac{\sinh^{-1}(\epsilon) }{\epsilon}-2 \right] 
 \label{def:le}
 \end{equation}

\noindent with $\sinh^{-1}$ the inverse of the hyperbolic sinus. We assume that the dome radius $b_d$ varies linearly with the radius of the jet $b_i$. An energy balance between the kinetic energy of mean flow   and the potential energy stored in the dome shows that $\epsilon=\xi /(2b_d) \sim Fr_i^2$, confirmed by the measurements of \cite{shy1995mixing}. For low Froude number $Fr_i < 1$, $\epsilon$ tends to zero and the length $l_e $ follows the asymptotic

 \begin{equation} 
\frac{l_e }{b_d}  \sim \epsilon^2 
 \end{equation}
 
 \noindent The model shows that the length $l$ tends to $b_d$  for $Fr_i\rightarrow 0$.   At large Froude number with $\epsilon \gg 1$, we obtain

 \begin{equation} 
\frac{l_e}{b_i}  \sim \epsilon
 \end{equation}
 
\noindent The two asymptotic behaviours  are given by the function $G(Fr_i)=l_e/b_i$  

\begin{equation}
G(Fr_i)=\frac{l_e}{b_i} \sim  \left \{\begin{array}{ll} \epsilon^2 \sim Fr_i^4 &\quad \hbox{for} \quad Fr_i\ll 1 \\ \\
\epsilon \sim Fr_i^2 & \quad \hbox{for} \quad  Fr_i\gg 1 
\end{array} \right.
\end{equation}

\noindent  Finally, the entrainment rate is given by 
\begin{equation}
E_i  \sim  \frac{l_e}{b_i} \frac{\sqrt{g' b_i}}{u_i} \sim Fr_i^{-1}  G(Fr)
\label{def:entra_m}
\end{equation}

\noindent hence

\begin{equation}
E_i\sim  \left \{\begin{array}{lll}  Fr_i^3 &\quad \hbox{for} \quad Fr_i\ll 1 \\ \\
 Fr_i& \quad \hbox{for} \quad  Fr_i\gg 1 
\end{array} \right.
\end{equation}

Our model recovers both asymptotic behaviours as a function of the Froude number (figure  \ref{fig:dome}(b)). We define the parametrization of the entrainment rate 

\begin{equation}
E_i= \alpha Fr_i^{-1}  G\left(\frac{\xi}{b_i}\right) \quad \hbox{with} \quad \frac{\xi}{b_i}=\beta Fr_i^2
\label{def:eq_Ei}
\end{equation}	 
 \noindent with $(\alpha,\beta)$ two real constants. The coefficient $\beta$ characterizes the crossover point $Fr_t=\beta^{-1/2}$, where the scaling law varies from $Fr_i^3$ to $Fr_i$.

\begin{figure} 
   
\centerline{ 
  \includegraphics[width=10cm,height=8cm]{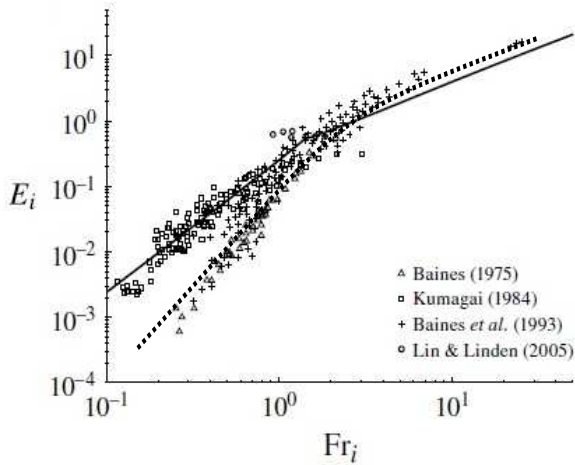} 
}
\caption{  Modified figure from \cite{shrinivas2014unconfined} (figure 12(a)). The dotted line corresponds to the scaling law defined by  equation  (\ref{def:eq_Ei} . {   The black line corresponds to the scaling law $E_i=0.24 Fr_i^2$ for small Froude numbers and $E_i=0.42 Fr_i$ at large Froude number, suggested  by \cite{shrinivas2014unconfined}. }\label{fig:erosion3}}  
\end{figure}

  	   	 We now  compare our  model to the  entrainment rate obtained in different processes of erosion (buoyant jets and  plumes). We use  figure 12(a)  from \cite{shrinivas2014unconfined}  containing the measurements of  \cite{baines1975entrainment} (triangles), \cite{kumagai1984turbulent} (squares), \cite{baines1993turbulent} (crosses) and \cite{lin2005entrainment} (circles).  
	   We have reported the curve defined by equation (\ref{def:eq_Ei}) with $(\alpha,\beta)=(1.5,0.45)$ in figure  \ref{fig:erosion3}  (dotted curve). {  These parameters are determined by the best fitting of the experimental points  of  \cite{baines1975entrainment} (triangles),  \cite{baines1993turbulent} (crosses) and \cite{lin2005entrainment} (circles). The value of the coefficient $\beta$ defines the crossover point occuring at $Fr_t=1.49$.} The agreement is good between the theory and the experimental data. To conclude, our model requires only one fitting amplitude parameter, here $\alpha$, to fit the data both at low and large Froude number, once we have fixed the crossover point (given by $\beta$), unlike former models which require  one parameter for each range.

\subsection{Data of Kumagai }
\label{sec:confinement}

{ 
    In  figure \ref{fig:erosion3},   the data of   \cite{kumagai1984turbulent} (upper branch)  display   a different trend relatively to the measurements of \cite{baines1975entrainment}, \cite{baines1993turbulent} and \cite{lin2005entrainment} (lower branch). The entrainment rate  is approximatively one order of magnitude larger than the other measurements. The exponent $n$ of the scaling law has been debated for small Froude numbers: \cite{kumagai1984turbulent} initially proposed  $n=3$ (dashed curve, figure \ref{fig:Kumagai}), whereas \cite{cardoso1993mixing} and \cite{shrinivas2014unconfined} suggested    $n=2$ (dotted line, figure \ref{fig:Kumagai}). The data being scattered, both models seem valid. Hence, the existence of the upper branch may be interpreted in two different ways. If $n$ is equal to $2$, two different physical mechanisms operate between the lower and the upper branch, as suggested in \cite{shrinivas2015confined}.  If  $n$ is equal to $3$, the mechanism is unchanged but    the pre-factors in the equation (\ref{def:eq_Ei}), $\alpha$ and $\beta$, vary with the conditions of the experiments, for instance because of confinement.
 
\begin{figure}  
\centerline{ \includegraphics[width=10cm,height=6cm]{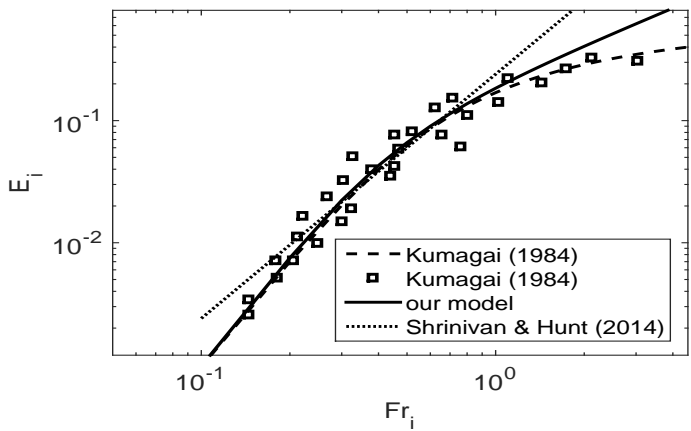}   }
\caption{Erosion rate $E_i(Fr_i)$   (black squares) from figure 12 in \cite{kumagai1984turbulent}. The dashed curve corresponds to the empirical model of  \cite{kumagai1984turbulent}. The dotted line is the scaling law from \cite{shrinivas2014unconfined} with $E_i=0.24 Fr_i^2$. The heavy curve corresponds to the model given by   (\ref{def:eq_Ei}) with $(\alpha,\beta)=(0.015,14)$, associated with a crossover point  at $Fr_t=0.27$. \label{fig:Kumagai}}  
\end{figure}
 
  
The data of   \cite{kumagai1984turbulent}  suggest that the transition between the behaviour at small and large Froude number occurs earlier compared to the lower branch.  We have reported  some values of the entrainment rate as a function of the Froude number (black squares) from figure 12 of \cite{kumagai1984turbulent} in  figure \ref{fig:Kumagai}. We observe a transition around $Fr_t \in [0.25-0.35]$ in the upper branch (figure \ref{fig:Kumagai})  whereas  the lower branch is characterized by $Fr_t = 1.49$ (figure \ref{fig:erosion3}). We recall that the coefficient $\beta$ in equation (\ref{def:eq_Ei}) {determines the value of the transition $Fr_t=\beta^{-1/2} $. Hence, the entrainment rates measured by \cite{kumagai1984turbulent} may differ from the ones of the lower branch because $\beta$ and   $\alpha$ are different.  Three models are reported in figure  \ref{fig:Kumagai}: the dashed   curve corresponds to the empirical model of \cite{kumagai1984turbulent}, the dotted line is the scaling law   $E_i=0.24 Fr_i^2$ from \cite{shrinivas2014unconfined} and the heavy black curve is given by the equation (\ref{def:eq_Ei}) with $(\alpha,\beta)=(0.015,14)$ with a   crossover point at  $Fr_t=0.27$. Our model follows the model of \cite{kumagai1984turbulent} up to $Fr_i\simeq 1$. Once again, the set of data is not sufficient to determine which model is the best, even if we recover the empirical model of \cite{kumagai1984turbulent}. However, our model shows that one possible origin of the upper branch is a {  modification} of the pre-factors between the experiments.

Our model is based on the deformation of the interface and when the interface is weakly deformed, the entrainment rate follows a scaling law $Fr_i^3$ for $Fr_i<Fr_t$. The coefficient $\beta$  determines the transition between a weakly  deformed interface, when the width of the dome is larger than its depth, and a strongly deformed interface, when the width   is significantly smaller than the depth. This transition should depend on the confinement, but also on the nature of the jet (buoyant or not) or the time-dependence of the erosion.  Hence, the scatter of the measurements may be explained by {  a modification} of the pre-factors $(\alpha,\beta)$, where $\beta$ characterizes the efficiency of the plume   to deform  the interface. But again, the physics of the erosion, dominated by breaking interfacial waves excited by the turbulent fluctuations and amplified by the mean flow, may remain the same.
 
 }
    
\section{Conclusion}
	
	In the present paper, we have investigated the mechanism of entrainment by a turbulent jet impinging on a density interface with moderate Reynolds number and moderate Froude number. In this regime, the vortices coming from the jet are not able to deform significantly the interface via  ballistic impacts, as commonly expected {  by previous model involving baroclinic turbulence inside the dome}.   Their role is reduced  to trigger interfacial  gravity waves. The main source of energy comes from the mean flow, which amplifies the perturbation of the interface by a mechanism {  of wave-induced stress. The sign of the Reynolds stress  changes abruptly  inside the dome,  where the phase velocity of the wave and the   mean velocity field are equal. These features suggest  the presence of  a resonance between the wave and the mean flow}. Wave amplification leads to wave-breaking in the vicinity of the border of the impinged region. The induced mixing is responsible for the irreversible erosion of the interface.  
	
	Based on these physical observations, we have introduced a scaling law, which varies continuously from the  $Fr_i^3$ to the $Fr_i$ power law from small to large Froude numbers, in agreement with   {  some of } previous experimental measurements. {   Our model offers an alternative   to \cite{shrinivas2015confined}: we suggest that the scatter of the measurements of the entrainment rate is rather due to a difference between the pre-factors of the scaling law rather than a modification of the  exponents.}

	Our measurements are performed for Reynolds numbers below the mixing transition, i.e. $Re<10^4$. It will be interesting to investigate the mechanism of erosion at low Froude number and  larger Reynolds numbers ($Re>10^4$), in order to verify if the present mechanism still operates or if others processes occur. {  For larger Reynolds numbers, the vortices impinging the stratification may be able   to generate baroclinic vortices and turbulence at the interface, as suggested by \cite{shy1995mixing}. The model based on baroclinic effect,  introduced by \cite{shrinivas2014unconfined} and \cite{shrinivas2015confined},  could be  then relevant}.

	\section*{Acknowledgements} 
The authors thank   three anonymous referees for their comments and  F. Duval (LIE, IRSN) and J.M. Ricaud (LIE, IRSN) for    stimulating discussions. They  acknowledge the support from the Institut de Radioprotection et de S\^uret\'e Nucl\'eaire and  region PACA (France) under the APEX program 2015 (Project S2URF).

	\bibliographystyle{jfm}
	\bibliography{jfm-instructions}

	\end{document}